\documentclass{aa}

\usepackage{graphicx}
\usepackage{txfonts}
\usepackage[colorlinks=true, linkcolor=blue, citecolor=blue]{hyperref}

\begin{document} 

\definecolor{coralpink}{rgb}{0., 0., 0.}

   \title{Particle-in-cell simulations of pulsar magnetospheres: transition between electrosphere and force-free regimes}

   \author{F. Cruz
          \inst{1,2}
          \and
          T. Grismayer
          \inst{1}
          \and
          A. Y. Chen
          \inst{3}
          \and
          A. Spitkovsky
          \inst{4}
          \and
          R. A. Fonseca
          \inst{1,5}
          \and
          L. O. Silva
          \inst{1}
          }

   \institute{GoLP/Instituto de Plasmas e Fus\~{a}o Nuclear, 
              Instituto Superior T\'{e}cnico, Universidade de Lisboa, 
              1049-001 Lisboa, Portugal\\
              \email{fabio.cruz@tecnico.ulisboa.pt}
         \and
              Inductiva Research Labs, Rua da Prata 80, 1100-420 Lisboa, Portugal
         \and
              Physics Department and McDonnell Center for the Space Sciences, Washington University in St. Louis, MO 63130, USA
         \and
              Department of Astrophysical Sciences, 
              Princeton University, Princeton, NJ 08544, USA
         \and
              DCTI/ISCTE Instituto Universitário de Lisboa,
              1649-026 Lisboa, Portugal
             }
             
    \date{Received \today}

 
  \abstract
   {}
   {Global particle-in-cell (PIC) simulations of pulsar magnetospheres are performed with a volume, surface and pair production-based plasma injection schemes to systematically investigate the transition between electrosphere and force-free pulsar magnetospheric regimes.}
   {A new extension of the PIC code OSIRIS to model pulsar magnetospheres using a two-dimensional axisymmetric spherical grid is presented. The sub-algorithms of the code and thorough benchmarks are presented in detail, including a new first-order current deposition scheme that conserves charge to machine precision.}
   {It is shown that all plasma injection schemes produce a range of magnetospheric regimes. Active solutions can be obtained with surface and volume injection schemes when using artificially large plasma injection rates, and with pair production-based plasma injection for sufficiently large separation between kinematic and pair production energy scales.}
   {}

   \keywords{pulsars -- electrosphere -- force-free magnetosphere -- particle-in-cell -- simulations}

   \maketitle
%

\section{Introduction}


Over the last decade, global kinetic simulations have been essential tools to understand the electrodynamics of pulsar magnetospheres. They have been used to study the organization of plasma currents in the vicinity of the neutron star~\citep{philippov_2015b, chen_2017, kalapotharakos_2018} and the acceleration of leptons~\citep{chen_2014, belyaev_2015, cerutti_2015, philippov_2014, philippov_2015a, brambilla_2018} and ions~\citep{guepin_2020} in the current sheets that develop beyond the light cylinder, leading to gamma-ray emission consistent with observations.

\textcolor{coralpink}{Particle-in-cell (PIC)~\citep{dawson_1962, dawson_1983, hockney_eastwood_1988, birdsall_langdon_1991} has been the main methodology used in global kinetic simulations of pulsar magnetospheres. PIC simulations reproduce with high fidelity the kinetic plasma phenomena relevant in pulsars, such as the evolution of highly non-thermal particle distributions or kinetic-scale fluctuations~\citep{touati_2022}. Recent extensions of the PIC method have also allowed the inclusion of Quantum Electrodynamics effects such as pair production~\citep{grismayer_2016, grismayer_2017} or general relativity corrections~\citep{philippov_2015b} relevant in pulsars.}

Due to the large disparity between kinetic and system scales in pulsars, PIC simulations typically employ a phenomenological description of the pair production processes responsible for filling the pulsar magnetosphere. Such description can be as simple as injecting plasma in a significant fraction the simulation domain~\citep{philippov_2014, belyaev_2015, kalapotharakos_2018, brambilla_2018}, limiting this injection to occur close to the stellar surface~\citep{cerutti_2015, hakobyan_2023}, or even considering heuristic pair production models~\citep{chen_2014, philippov_2015a, philippov_2015b, chen_2020a, guepin_2020, bransgrove_2022}. 

Depending on the details of the injection and/or pair production model, the global asymptotic magnetospheric topology varies quite significantly: in some cases, the system auto-regulates to a fully charge-separated configuration (also called electrosphere) that does not produce a Poynting flux, whereas in other cases the magnetosphere converges to a force-free regime~\citep{philippov_2014, chen_2014, cerutti_2015, guepin_2020, hakobyan_2023}. While this range of solutions has been identified in several works, a systematic study has not been performed to compare volume, surface and pair production-based injection schemes.

\textcolor{coralpink}{In this work, we perform two-dimensional axisymmetric global simulations of pulsar magnetospheres \textcolor{coralpink}{with three different pair injection schemes: over large volumes of the magnetosphere, from the stellar surface only and using a prescription model for pair production. We use these simulations to} to systematically characterize the obtained magnetospheric solutions as a function of the injection and/or pair production model parameters. We show that all plasma sources produce near force-free solutions in the regime of large plasma supply and inactive electrosphere solutions with small plasma supply. All plasma sources also allow a transitional regime with sub-force-free surface Poynting flux and wide equatorial current sheets.}

\textcolor{coralpink}{The simulations presented in this work are performed with a recent extension of the PIC code OSIRIS~\citep{fonseca_2002, fonseca_2008} developed for magnetospheric models of compact objects, presented also in this work for completeness.}


This paper is organized as follows. In Sect.~\ref{sec:numerical_tool}, we describe the set of numerical techniques used to generalize the PIC method to perform two-dimensional \textcolor{coralpink}{axisymmetric} global kinetic simulations of pulsar magnetospheres with OSIRIS: the adopted discretization of the spatial domain is presented in Sect.~\ref{sec:spherical_grid} and the numerical schemes used to advance the field and particle equations and the corresponding boundary conditions are detailed in Sects.~\ref{sec:spherical_solver} and \ref{sec:spherical_pusher}. A new charge-conserving current deposition scheme is presented in Sect.~\ref{sec:spherical_current}, and the typical \textcolor{coralpink}{scales and} normalizations \textcolor{coralpink}{adopted in the code} are presented in Sect.~\ref{sec:spherical_normalizations}. In Sect.~\ref{sec:global}, we present simulations with volume (Sect.~\ref{sec:global_vol}), surface (Sect.~\ref{sec:global_sur}) and pair production-based (Sect.~\ref{sec:global_pp}) plasma injection. Our conclusions are presented in Sect.~\ref{sec:conclusions}.

\section{Numerical tool}
\label{sec:numerical_tool}

\subsection{Discretization and spatial grid}
\label{sec:spherical_grid}

The numerical tool presented in this work aims to model the global plasma environment surrounding neutron stars, \textit{i.e.}, the spatial volume \textcolor{coralpink}{between the stellar surface and a few light cylinder radii above it}. We describe this system in spherical coordinates, with the radial coordinate $r$ measured from the center of the neutron star and the polar angle $\theta$ measured from the star's rotation axis $\boldsymbol{\Omega}$. We assume that $\boldsymbol{\Omega}$ is either parallel or anti-parallel to the star's magnetic axis $\boldsymbol{\mu}$, such that we can assume axisymmetry about $\boldsymbol{\Omega}$, \textit{i.e.}, derivatives with respect to the azimuthal angle $\phi$ can be dropped, $\partial / \partial \phi = 0$.

\textcolor{coralpink}{Similarly to \citet{chen_2014, cerutti_2015}}, we discretize the simulation domain $r \in [r_\mathrm{min}, r_\mathrm{max}]$, $\theta \in [0, \pi]$ in a grid with $N_r \times N_\theta$ cells. We adopt a regular grid spacing in $\theta$, $\Delta \theta = \pi / (N_\theta + 1)$, and in $\log r$. The latter choice allows for a grid spacing that monotonically increases with $r$. In pulsar magnetosphere simulations, this choice favors the resolution of shorter spatial scales close to the stellar surface, where denser plasmas are expected, and relaxes it far from the neutron star, where it is less needed. The discretization in the radial direction can be formally written as
\begin{equation}
\log r_n = \log r_\mathrm{min} + (n-1) \Delta \ , \ \ n = 1, 2, ..., N_{r}+1 \ ,
\label{eq:discretization_logr}
\end{equation}
with $\Delta \equiv \log(r_\mathrm{max} / r_\mathrm{min}) / N_{r}$. Equation~\eqref{eq:discretization_logr} can be manipulated to write the useful relation $r_n = r_\mathrm{min} \delta^{n-1}$, where $\delta \equiv (r_\mathrm{max} / r_\mathrm{min})^{1/N_r}$ is a parameter that combines all properties of the radial axis.

\begin{figure*}
\centering
\includegraphics[width=5.6in]{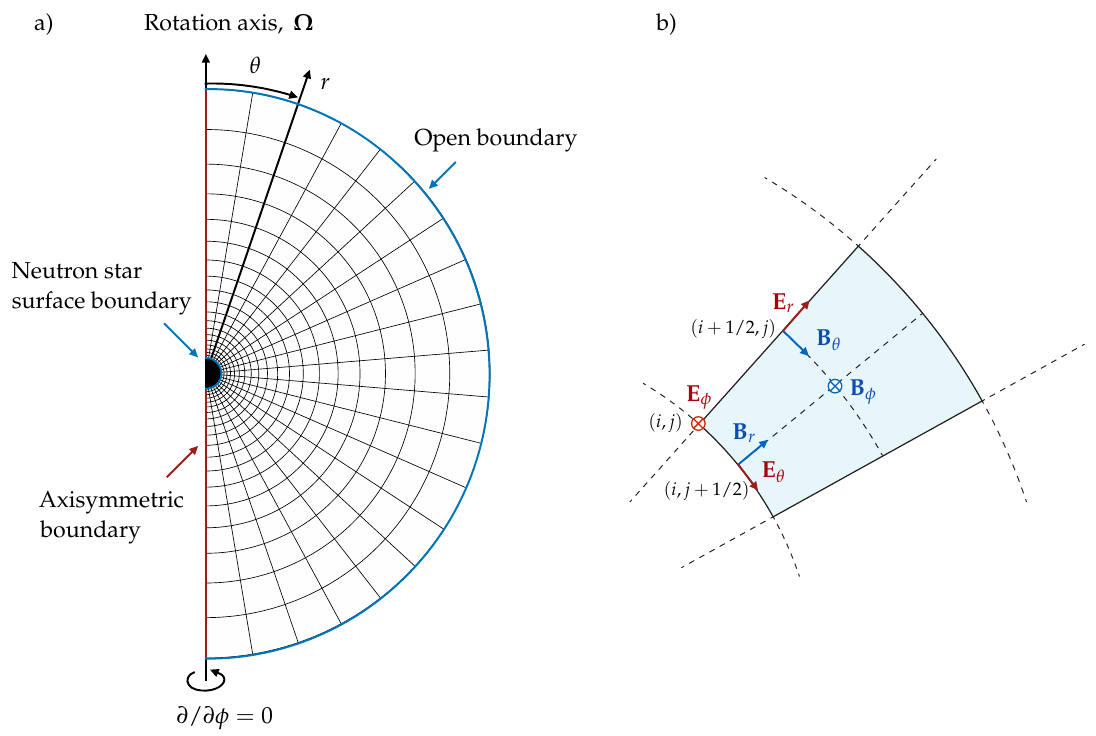}
\caption{Schematic representation of spherical PIC grid: a) shows the grid layout and identifies the coordinate system and boundary types, b) shows the grid cell's edges where each field component is defined.}
\label{fig:spherical_grid}
\end{figure*}

A schematic representation of the grid used to discretize a typical simulation domain in illustrated in Fig.~\ref{fig:spherical_grid}a. The edges of grid cells are shown in black lines, and domain boundaries are highlighted in blue and dark red. The lower radial boundary coincides with the stellar surface, $r_\mathrm{min} = r_*$, whereas the upper radial boundary is at $r_\mathrm{max} \sim$~tens of $r_*$, and acts as an open boundary. The $\theta = {0, \pi}$ boundaries enforce axisymmetry, effectively serving as reflecting boundaries. More details about these boundaries are provided in Sects.~\ref{sec:spherical_solver} and \ref{sec:spherical_pusher}.

In Fig.~\ref{fig:spherical_grid}b, we show a schematic representation of a typical grid cell, that we label with indices $(i,j)$ in the radial and polar directions, respectively. Cell boundaries are drawn in solid black lines, and auxiliary lines are drawn in dashed black lines. The positions where the electric and magnetic field components are defined are indicated in dark red and blue. Half integer indices $i+1/2$ and $j+1/2$ indicate positions defined as $r_{i+1/2} \equiv (r_i + r_{i+1})/2$ and $\theta_{j+1/2} \equiv (\theta_j + \theta_{j+1})/2$, respectively. The grid illustrated in Fig.~\ref{fig:spherical_grid} presents two key differences with respect to a typical Cartesian grid: a) its cells have curvilinear boundaries and b) their shape and volume change across the grid. These conditions make each step of the PIC method in spherical coordinates more \textcolor{coralpink}{challenging, requiring conversions between coordinate systems in the particle pusher and adjustments in the current deposition scheme to accomodate particle shrinking/expansion in each time step}. We explore these \textcolor{coralpink}{challenges} and workarounds in Sects.~\ref{sec:spherical_solver}, \ref{sec:spherical_pusher} and \ref{sec:spherical_current}.

\subsection{Electromagnetic field solver}
\label{sec:spherical_solver}

Electric and magnetic field components are defined in the edges of the staggered grid cells indicated in Fig.~\ref{fig:spherical_grid}b. This definition is analogous to that used in traditional Cartesian grids, and allows the use of the Yee algorithm~\citep{yee_1966} to advance the electric and magnetic field in time via Maxwell's equations,
\begin{align}
& \mathbf{B}^{n+1/2} = \mathbf{B}^{n-1/2} = - c \Delta t (\nabla \times \mathbf{E})^{n} \ , \label{eq:faraday_disc_time} \\
& \mathbf{E}^{n+1} =  \mathbf{E}^{n} + c \Delta t (\nabla \times \mathbf{B})^{n+1/2} - 4 \pi \Delta t \mathbf{j}^{n+1/2} \ \label{eq:ampere_disc_time} ,
\end{align}
where quantities with integer/half integer superscripts are defined in integer/half integer times and $\Delta t$ is the time step.

Here we adopt the same methodology as~\citet{cerutti_2015, belyaev_2015b} and use an integral form of Maxwell's equations that avoids divergences on the polar boundaries. This integral form is obtained by using Stokes' theorem to evaluate the curl of electric and magnetic fields in a given cell as
\begin{equation}
(\nabla \times \mathbf{E})_\mathrm{cell} = \left( \oint_{\mathcal{C}_\mathrm{cell}} \mathbf{E} \cdot \mathrm{d}\mathcal{C}_\mathrm{cell} \right) / \mathcal{S}_\mathrm{cell} \ ,
\label{eq:stokes}
\end{equation}
where $\mathcal{C}_\mathrm{cell}$ is the contour defining the edge of that cell, $\mathcal{S}_\mathrm{cell}$ is the corresponding area, and the closed integral and dot product have the usual definition of Stokes' theorem. The cell label and corresponding integrations in Eq.~\eqref{eq:stokes} change according to the field component under consideration. For instance, we can write the radial component of $\nabla \times \mathbf{E}$ as
\begin{equation}
{(\nabla \times \mathbf{E})_r}_{(i, j+1/2)} = \frac{\sin \theta_{j+1} {E_\phi}_{(i,j+1)} - \sin \theta_{j} {E_\phi}_{(i,j)}}{r_i (\cos \theta_j - \cos \theta_{j+1})} \ .
\label{eq:stokes_er}
\end{equation}
This expression is derived by noting that, according to Eq.~\eqref{eq:faraday_disc_time}, $(\nabla \times \mathbf{E})_r$ should be defined in the same position as $\mathbf{B}_r$, \textit{i.e.}, at cell indices $(i, j+1/2)$. This defines the integration surface relevant to Stokes' theorem as $r = r_i$, $\theta \in [\theta_j, \theta_{j+1}]$. The numerator and denominator in Eq.~\eqref{eq:stokes} then read respectively $2 \pi (r_i \sin \theta_{j+1} {E_\phi}_{(i,j+1)} - r_i \sin \theta_j {E_\phi}_{(i,j)})$ and $2 \pi r_i^2 (\cos \theta_j - \cos \theta_{j+1})$, where the $2 \pi$ factor comes from the integration along $\phi$. A similar calculation can be performed for all other components~\citep{cerutti_2015}.

We note that at the simulation boundaries ($i = \{1, N_r+1\}$, $j = \{1, N_\theta+1\}$), the integration regions are adapted to fit within the domain. For example, the $\theta$ integration is changed to $\theta \in [0, \theta_{1+1/2}]$ and $\theta \in [\theta_{N_\theta+1/2}, \pi]$ at the $\theta = 0$ and $\theta = \pi$ boundaries, respectively. We also apply special rules to the field components at the boundaries, e.g. in the polar boundaries we enforce the axisymmetry conditions ${\mathbf{E}_\phi}_{(i,1)} = {\mathbf{E}_\phi}_{(i,N_\theta+1)} = 0$ and ${\mathbf{B}_\theta}_{(i+1/2,1)} = {\mathbf{B}_\theta}_{(i+1/2,N_\theta+1)} = 0$. The inner radial boundary acts generally as a rotating conductor mimicking the stellar surface, whereas the outer boundary acts as a first-order standard Mur open boundary condition~\citep{mur_1981}\textcolor{coralpink}{, \textit{i.e.}, a perfect absorber of perturbations propagating perpendicularly to the boundary}. We have also implemented static conductor boundary conditions for both inner and outer radial boundaries, that enforce tangent (normal) electric (magnetic) field components to be null, \textit{i.e.}, ${\mathbf{E}_\phi}_{(1,j)} = {\mathbf{E}_\phi}_{(N_r+1,j)} = 0$, ${\mathbf{E}_\theta}_{(1,j+1/2)} = {\mathbf{E}_\theta}_{(N_r+1,j+1/2)} = 0$ and ${\mathbf{B}_r}_{(1,j+1/2)} = {\mathbf{B}_r}_{(N_r+1,j+1/2)} = 0$.

We have benchmarked our field solver implementation by studying stationary electromagnetic TM modes between two spherical static conductors~\citep{jackson_1975}. We have verified that the solution obtained numerically is in excellent agreement with the analytical solution of Maxwell's equations for these modes, as well as with the detailed discussion about a similar solver in \citet{belyaev_2015b}. 

\subsection{Particle pusher}
\label{sec:spherical_pusher}

Particle position and momentum components are updated in Cartesian coordinates with either the Boris~\citep{boris_1970, birdsall_langdon_1991} or Vay~\citep{vay_2008} pushers\textcolor{coralpink}{, although other pushers are also compatible with the remaining modified sub-algorithms of PIC presented in this work}. In each time step, a particle push is done as follows: first, the electric and magnetic fields are interpolated from the edges of the corresponding grid cell to the particle position $\mathbf{x}_p^n \equiv (r_p, \theta_p)$, an operation that we write schematically as $(\mathbf{E}^n_{(i,j)}, \mathbf{B}^n_{(i,j)}) \to (\mathbf{E}^n_p, \mathbf{B}^n_p)$. This interpolation is done using a area/volume weighting scheme. For example, the toroidal component of the electric field can be written as
\begin{equation}
{\mathbf{E}_\phi}_p = \sum_{i' = i, i+1} \sum_{j' = j, j+1} {f_r}_{i'} {f_\theta}_{j'} {\mathbf{E}_\phi}_{(i',j')} \ ,
\end{equation}
with
\begin{align}
{f_r}_{i} &= 1 - {f_r}_{i+1} = \frac{r_p^3 - r_i^3}{r_{i+1}^3 - r_{i}^3} \ , \nonumber \\
{f_\theta}_{j} &= 1 - {f_\theta}_{j+1} = \frac{\cos \theta_j - \cos \theta_p}{\cos \theta_j - \cos \theta_{j+1}} \ . \nonumber
\end{align}
After the interpolation, the field components are converted from spherical to Cartesian coordinates, $(\mathbf{E}^n_p, \mathbf{B}^n_p) \to (\mathbf{E}^n_{p, \mathrm{C}}, \mathbf{B}^n_{p, \mathrm{C}})$, a \textcolor{coralpink}{calculation} that depends on the particle position at time $t^n$, $\mathbf{x}^n$. Finally, the particle momentum and position are updated in time, $\mathbf{u}^{n-1/2} \equiv \mathbf{p}^{n-1/2} / m_e c \to \mathbf{u}^{n+1/2} \equiv \mathbf{p}^{n+1/2} / m_e c$ and $\mathbf{x}^{n} \to \mathbf{x}^{n+1}$ respectively. Choosing to advance position and momentum components in Cartesian coordinates guarantees that we are solving the simplest possible equations of motion and also allows for an easy integration with other modules in OSIRIS, such as those accounting for classical radiation reaction losses~\citep{vranic_2016} and QED effects~\citep{grismayer_2016, grismayer_2017}. We note that advancing the particle position in $(x,y,z)$ does not introduce any asymmetry in the azimuthal direction $\phi$; in fact, each macro-particle in our simulation represents a charged ring with azimuthal symmetry and $\phi$ is never used throughout the rest of the numerical scheme.

\begin{figure*}
\centering
\includegraphics[width=4.0in]{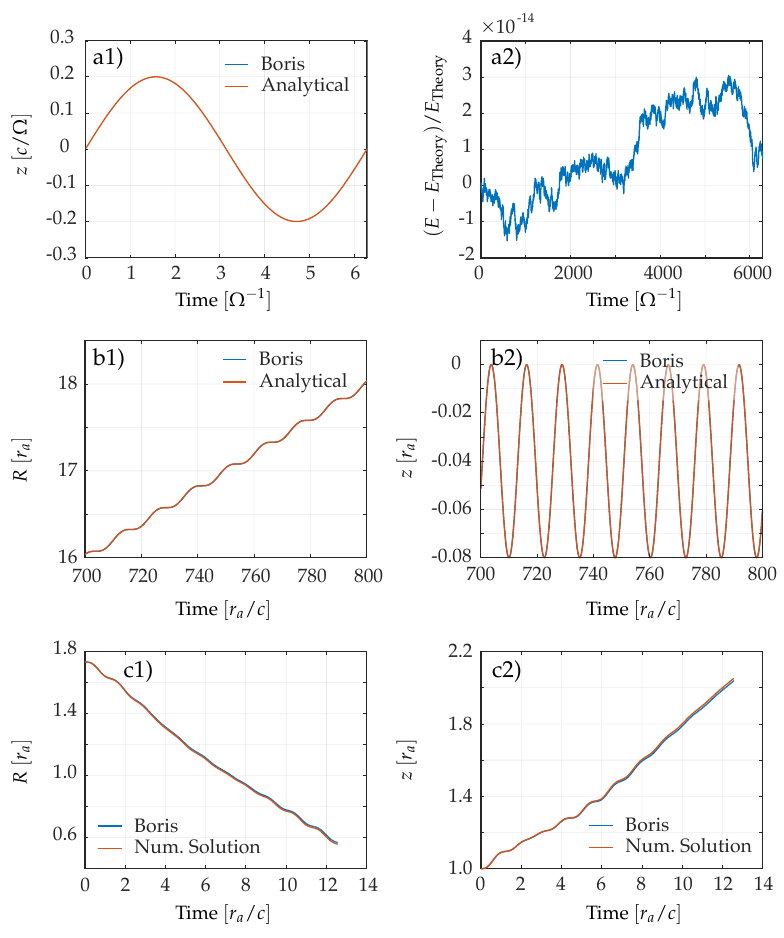}
\caption{Particle pusher benchmarks corresponding to particle motions in a1-2) a uniform azimuthal magnetic field, b1-2) crossed constant magnetic and electric fields and c1-2) the time-varying electric and magnetic field components of TM modes.}
\label{fig:pusher_benchmarks}
\end{figure*}

We have tested our implementation of the particle pushers in a large set of background electric and/or magnetic field configurations. In Fig.~\ref{fig:pusher_benchmarks}, we show results from a relevant subset of these configurations, namely a particle moving in a) a uniform azimuthal magnetic field, b) crossed constant magnetic and electric fields and c) the time-varying electric and magnetic field components of the TM modes described in the electromagnetic field solver benchmark presented in Sect.~\ref{sec:spherical_solver}. For all these cases, we show a comparison between the solutions obtained with the Boris pusher and analytical or other numerical solutions. We obtain an excellent agreement between the results of the Boris pusher and the reference analytical/numerical curves. Solutions obtained with the Vay pusher show a similar agreement with the reference curves. In Fig.~\ref{fig:pusher_benchmarks}a2, we represent the temporal evolution of the particle energy for over $\sim$~1000 periods, showing that it is conserved to machine precision. We note that in all these benchmarks, the only electromagnetic fields were those either imposed externally or calculated with the field solver, \textit{i.e.}, they do not include the fields self-consistently created due to particle motion via plasma currents. 

\subsection{Current deposition}
\label{sec:spherical_current}

A current deposition algorithm computes the current density $\mathbf{j}$ on the edges of grid cells as the positions and momenta of particles are updated. A trivial choice is to compute this current as the sum over the macro-particles of the product of their charge density and instantaneous velocity. However, such algorithm in general does not satisfy the continuity equation~\citep{villasenor_1992, esirkepov_2001},
\begin{equation}
\frac{\partial \rho}{\partial t} + \nabla \cdot \mathbf{j} = 0 \ ,
\label{eq:continuity_code}
\end{equation}
where $\rho$ is the total plasma density. Solving Eq.~\eqref{eq:continuity_code} ensures also that Gauss' law, written as
\begin{equation}
\nabla \cdot \mathbf{E} = 4 \pi \rho \ ,
\label{eq:gauss_code}
\end{equation}
is satisfied. Finding a current deposition algorithm that satisfies Eq.~\eqref{eq:continuity_code}, and consequently Eq.~\eqref{eq:gauss_code}, \textit{i.e.}, a charge-conserving current deposition algorithm, is one of the \textcolor{coralpink}{key} challenges in PIC codes. For Cartesian grids, there is a well established method for any interpolation order proposed in \citet{esirkepov_2001}. However, for non-uniform spherical grids, this challenge is more substantial, as grid cells (and particle shapes, that we shall define below) change across the grid. Other codes adopting such grids~\citep{chen_2014, cerutti_2015, belyaev_2015b, chen_2017} usually do not \textcolor{coralpink}{seem to} include charge-conserving current deposition algorithms, and adopt instead numerical schemes to enforce the validity of Eq.~\eqref{eq:gauss_code}\textcolor{coralpink}{, e.g. Poisson solvers}.

\begin{figure*}[t]
\centering
\includegraphics[width=5.2in]{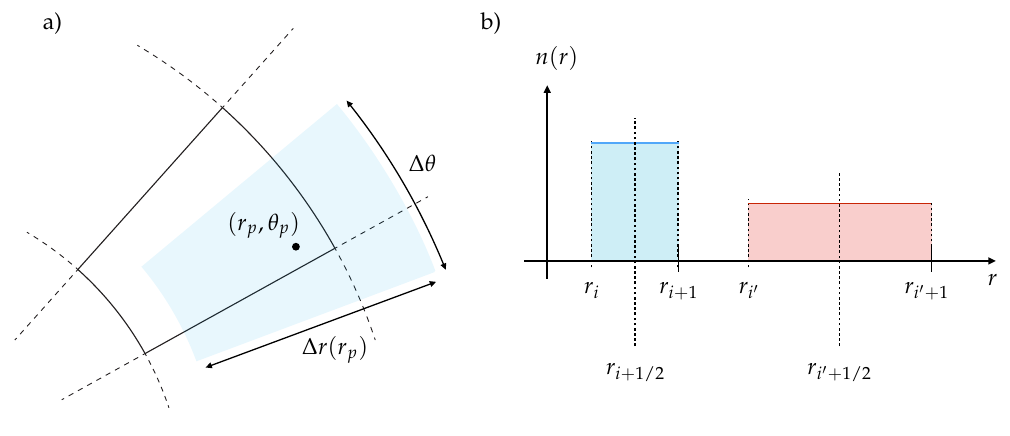}
\caption{Schematic representation of a) the spherical particle shape and b) the variation of its flat-top density value with the radial coordinate. The blue shaded region in a) represents the particle shape and identifies its widths in the radial and polar directions.}
\label{fig:spherical_shape}
\end{figure*}

Here, we propose a new current deposition scheme that conserves charge to machine precision in the non-uniform grid defined in section~\ref{sec:spherical_grid}. We start by defining the volume occupied by a macro-particle centered at $(r_p, \theta_p)$. The function that defines this volume is usually called the particle shape, $S(r, \theta, r_p, \theta_p)$. Before writing the exact form of $S$, let us define some of its important properties, that we illustrate schematically in Fig.~\ref{fig:spherical_shape}. First, the particle shape should only coincide with the shape of the cell in which its center is located, labeled with indices $(i,j)$, when and only when $(r_p, \theta_p) = (r_{i+1/2}, \theta_{j+1/2})$. Since the grid spacing in the radial direction is a function of $r$, the particle width in this direction should also be a function of $r_p$, \textit{i.e.}, $\Delta r \equiv \Delta r (r_p)$. Furthermore, the charge density associated with each macro-particle should also be a function of $r_p$. More specifically, the charge density should decrease with $r_p$ to compensate the corresponding increase in volume of the macro-particle, such that its total charge remains constant.

\begin{figure*}[t]
\centering
\includegraphics[width=4.0in]{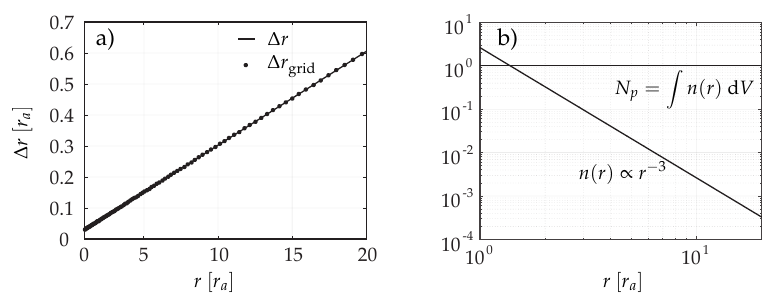}
\caption{Particle shape properties: a) radial width and b) density and real particle number.}
\label{fig:spherical_shape_dep}
\end{figure*}

Defining the number of real particles in a macro-particle as $N_p$, we formally wish to find a waterbag-like particle number density $n(r)$ such that
\begin{equation}
\int_{V_i} n(r_{i+1/2}) \ \mathrm{d}V_i = \int_{V_{i'}} n(r_{i'+1/2}) \ \mathrm{d}V_{i'} = N_p \ ,
\label{eq:shape_r_def}
\end{equation}
where $V_{i, i'}$ are the volumes of cells with radial labels $i, i'$ (see Figure~\ref{fig:spherical_shape} b)). \textcolor{coralpink}{For simplicity, we} assume that the particle density is only a function of $r$, and generalize it later to include the natural dependence in $\theta$ as well. Assuming that $n(r_{i+1/2})$ is constant within cell $i$, we can solve Eq.~\eqref{eq:shape_r_def} to obtain
\begin{equation}
n(r_{i+1/2}) = \frac{3 N_p}{4\pi} \frac{1}{r_{i+1}^3 - r_i^3} = \frac{3N_p}{32\pi} \frac{\left(\delta + 1\right)^3}{\delta^3 - 1}  \frac{1}{r_{i+1/2}^3} \ ,
\label{eq:shape_r_nr}
\end{equation}
where we have used the relation $r_{i+1/2} = r_i (1 + \delta) / 2 = r_{i+1} ( 1 + \delta^{-1}) / 2$. We note that Eq.~\eqref{eq:shape_r_nr} defines $n(r)$ for any $r_{i+1/2}$, but \textcolor{coralpink}{not for} $r \ne r_{i+1/2}$. We choose to take the continuous limit of $n(r_{i+1/2})$ for an arbitrary radius, \textcolor{coralpink}{replacing} $r_{i+1/2}$ for an arbitrary $r_p$, \textcolor{coralpink}{\textit{i.e.}},
\begin{equation}
n(r_p) = \frac{3N_p}{32\pi} \frac{\left(\delta + 1\right)^3}{\delta^3 - 1} \frac{1}{r_p^3} \ .
\label{eq:shape_r_nr_part}
\end{equation}
Eq.~\eqref{eq:shape_r_nr_part} ensures that $n(r)$ satisfies exactly Eq.~\eqref{eq:shape_r_def} when $r_p = r_{i+1/2}$ and that the particle shape is a smooth function of $r_p$. The particle width $\Delta r(r_p)$ is determined in a similar manner; first, we express the grid spacing in terms of $r_{i+1/2}$, $\Delta r_i = r_{i+1} - r_i = 2 r_{i+1/2} (\delta-1)/(\delta+1)$, and we extend this definition to an arbitrary radius $r_p$,
\begin{equation}
\Delta r (r_p) = 2 r_p \frac{\delta-1}{\delta+1} \ .
\label{eq:shape_r_dr_part}
\end{equation}
This quantity is represented for a typical grid in Fig.~\ref{fig:spherical_shape_dep}a, together with the grid spacing $\Delta r_i$. As expected, both quantities match exactly when $r = r_{i+1/2}$, and $\Delta r$ is a smooth function of $r$. Equations~\eqref{eq:shape_r_nr_part} and \eqref{eq:shape_r_dr_part} ensure that the conservation law expressed in Eq.~\eqref{eq:shape_r_def} can be extended to any radius, which is shown in Fig.~\ref{fig:spherical_shape_dep}b.

The general particle shape $S$ can be inferred from this discussion, and in particular from Eq.~\eqref{eq:shape_r_nr_part}. It reads
\begin{align}
S(r, \theta, r_p, \theta_p) &= \frac{3}{16\pi} \frac{\left(\delta + 1\right)^3}{\delta^3 - 1} \frac{1}{r_p^3} b_0 \left( \frac{r - r_p}{\Delta r (r_p)} \right) \times \nonumber \\
& \times \frac{1}{\cos (\theta_p - \Delta \theta / 2) - \cos (\theta_p + \Delta \theta / 2)} b_0 \left( \frac{\theta - \theta_p}{\Delta \theta} \right) \ ,
\label{eq:shape_gen}
\end{align}
where $b_0(x)$ is the zeroth order b-spline function, defined as $b_0(x) = 1$ if $|x| < 0.5$ and $0$ otherwise. Note that Eq.~\eqref{eq:shape_gen} generalizes the particle shape to a two-dimensional $(r, \theta)$ grid, hence the $\cos(\theta_p \pm \Delta \theta)$ terms resulting from the integral in Eq.~\eqref{eq:shape_r_def}. With the shape function in Eq.~\eqref{eq:shape_gen}, we can compute the charge density at any point $(r, \theta)$ due to the presence of a macro-particle with $N_p$ real particles of charge $q_p$ and coordinates $(r_p, \theta_p)$ as $\rho_p(r, \theta, r_p, \theta_p) = q_p N_p S(r, \theta, r_p, \theta_p)$. The charge density at cell edges is defined resorting to the area/volume weighting technique described in Sect.~\ref{sec:spherical_pusher}, and can be formally derived as
\begin{align}
\rho_{(i,j)}(r_p, \theta_p) &= \frac{\int_{V_{i,j}} \rho_s (r, \theta, r_p, \theta_p) \ \mathrm{d}V_{i,j}}{V_{i,j}} = \nonumber \\
&= \dfrac{q_p N_p \dfrac{3}{16\pi} \dfrac{\left(\delta + 1\right)^3}{\delta^3 - 1}}{(r_{i+1/2}^3 - r_{i-1/2}^3) (\cos \theta_{j-1/2} - \cos \theta_{j+1/2})} \times \nonumber \\
& \times \Bigg[ \frac{r_>^3 - r_<^3}{r_p^3} \Bigg] \Bigg[ \frac{\cos (\theta_p - \Delta \theta / 2) - \cos \theta_{j+1/2}}{\cos (\theta_p - \Delta \theta / 2) - \cos (\theta_p + \Delta \theta / 2)} \Bigg] \ .
\label{eq:shape_dep}
\end{align}
We note that the special integration limits $r_> = \min ( r_p + \Delta r (r_p) / 2, r_{i+1/2})$ and $r_< = \max ( r_p - \Delta r (r_p) / 2, r_{i-1/2})$ result from the subtlety that the particle radial width is a function of the particle radial coordinate, $r_p$. The expressions in square brackets are often referred to as the weighting functions in PIC current deposition algorithms.

\begin{figure*}[t]
\centering
\includegraphics[width=5.4in]{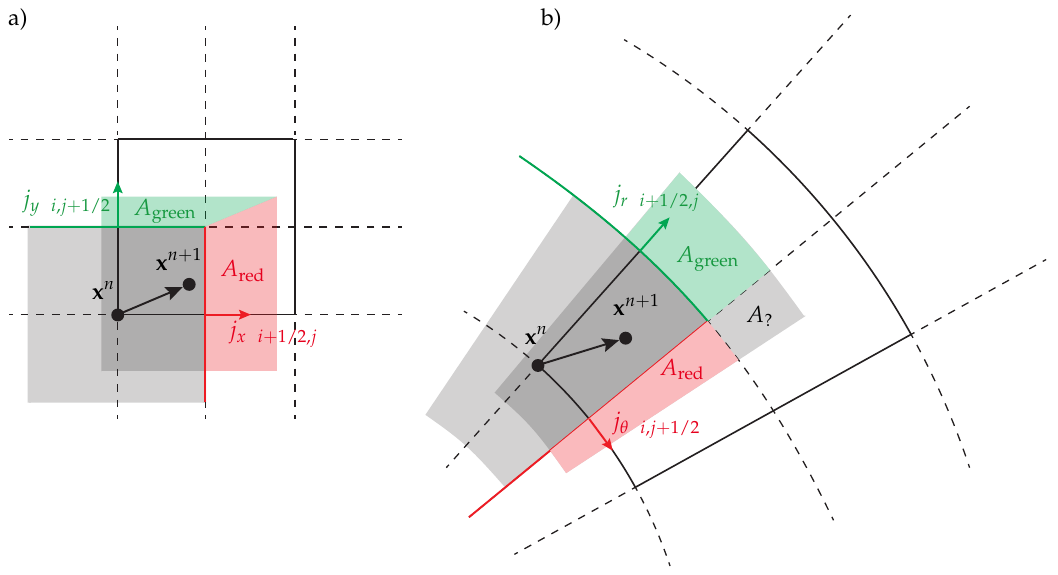}
\caption{Schematic representation of the current deposition algorithm in a) Cartesian and b) spherical coordinates (see text for details).}
\label{fig:spherical_deposit}
\end{figure*}

The particle shape in Eq.~\eqref{eq:shape_gen} and the deposition rule in Eq.~\eqref{eq:shape_dep} are \textcolor{coralpink}{the} key ingredients in our charge-conserving current deposition scheme. This scheme is inspired by the seminal work \textcolor{coralpink}{of}~\citet{villasenor_1992} (hereafter VB), that presented a scheme that predecessed the widely used method of~\citet{esirkepov_2001} for PIC current deposition in Cartesian grids. The VB method is schematically represented in Fig.~\ref{fig:spherical_deposit}a. VB proposed that the current density $\mathbf{j}$ should be computed directly by inverting the continuity equation, thus enforcing by construction that it is satisfied. In practice, when a particle is pushed in time from a position $\mathbf{x}^n$ to a position $\mathbf{x}^{n+1}$, part of its shape crosses the boundaries over which the current density is defined in the Cartesian PIC grid. These boundaries, and the exact locations where each of the components of $\mathbf{j}$ are defined are shown in Fig.~\ref{fig:spherical_deposit}a in green and red lines and arrows, respectively. VB recognized that we can simply compute the different current density components by evaluating the fraction of charge density carried by each macro-particle that crosses the boundaries identified in green and red. For a Cartesian grid, this fraction can be computed geometrically as the ratio between the areas $A_\mathrm{green}$ and $A_\mathrm{red}$ and the total area corresponding to the particle shape, $A_\mathrm{total}$. This calculation is simple in Cartesian grids because the particle shape does not change across the grid, which allows us to label which parts of the colored area at $x > x_{i+1/2}$ and $y > y_{j+1/2}$ crossed each of the green or red lines. In a spherical grid, this condition is not met, and the calculation becomes more involved. 

A schematic representation of the method equivalent to VB in a spherical grid is shown in Fig.~\ref{fig:spherical_deposit}b, where same rationale described above is easily applied except for the determination of the area identified with $A_\mathrm{?}$. Because the particle expands during its motion from $\mathbf{x}^n$ to $\mathbf{x}^{n+1}$, it is not trivial to determine which fraction of $A_\mathrm{?}$ should be combined with $A_\mathrm{green}$ ($A_\mathrm{red}$) to compute ${j_r}_{(i+1/2,j)}$ (${j_\theta}_{(i,j+1/2)}$). We circumvent this issue by generalizing the geometrical interpretation of $\nabla \cdot \mathbf{j}$ proposed by VB. They suggested that the total current divergence can be split as $\nabla \cdot \mathbf{j} = (\nabla \cdot \mathbf{j})_x + (\nabla \cdot \mathbf{j})_y$ in a Cartesian grid, with $(\nabla \cdot \mathbf{j})_y \propto A_\mathrm{green} / A_\mathrm{total}$ and $(\nabla \cdot \mathbf{j})_x \propto A_\mathrm{red} / A_\mathrm{total}$, and that these terms could be computed directly by evaluating $- \partial \rho_{(i,j)} / \partial t$ assuming that the particle moves purely along the corresponding direction at an average position along the orthogonal direction. Formally, this is expressed as
\begin{align}
{(\nabla \cdot \mathbf{j})_x}_{(i,j)} &= - \left. \frac{\partial \rho_{(i,j)}}{\partial t} \right|_{x^n, \bar{y}}^{x^{n+1}, \bar{y}}  = \frac{\rho_{(i,j)} (x^{n+1}, \bar{y}) -  \rho_{(i,j)} (x^{n}, \bar{y})}{\Delta t}\ , \label{eq:div_jx} \\
{(\nabla \cdot \mathbf{j})_y}_{(i,j)} &= - \left. \frac{\partial \rho_{(i,j)}}{\partial t} \right|_{\bar{x}, y^n}^{\bar{x}, y^{n+1}}  = \frac{\rho_{(i,j)} (\bar{x}, y^{n+1}) -  \rho_{(i,j)} (\bar{x}, y^{n})}{\Delta t}\ , \label{eq:div_jy}
\end{align}
where $\bar{x} = (x^{n+1} + x^n) / 2$ and $\bar{y} = (y^{n+1} + y^n) / 2$. From Eqs.~\eqref{eq:div_jx} and \eqref{eq:div_jy}, we can express the divergence operators using finite differences and obtain ${j_x}_{(i+1/2,j)}$ and ${j_y}_{(i,j+1/2)}$. This approach can be generalized to spherical coordinates, \textit{i.e.}, we can write $\nabla \cdot \mathbf{j} = (\nabla \cdot \mathbf{j})_r + (\nabla \cdot \mathbf{j})_\theta$. However, because the particle shape changes continuously in the radial direction, $(\nabla \cdot \mathbf{j})_\theta$ cannot be computed assuming that the particle moves purely along the polar direction with $\bar{r} = (r^{n+1} + r^n) / 2$. Instead, we proceed as follows: first, we compute $\nabla \cdot \mathbf{j}$ and $(\nabla \cdot \mathbf{j})_r$ using
\begin{align}
{(\nabla \cdot \mathbf{j})}_{(i,j)} &= - \left. \frac{\partial \rho_{(i,j)}}{\partial t} \right|_{r^n, \theta^n}^{r^{n+1}, \theta^{n+1}}  = \frac{\rho_{(i,j)} (r^{n+1}, \theta^{n+1}) -  \rho_{(i,j)} (r^{n}, \theta^n)}{\Delta t}\ , \label{eq:div_jtot} \\
{(\nabla \cdot \mathbf{j})_r}_{(i,j)} &= - \left. \frac{\partial \rho_{(i,j)}}{\partial t} \right|_{r^n, \bar{\theta}}^{r^{n+1}, \bar{\theta}}  = \frac{\rho_{(i,j)} (r^{n+1}, \bar{\theta}) -  \rho_{(i,j)} (r^{n}, \bar{\theta})}{\Delta t}\ , \label{eq:div_jr}
\end{align}
where $\bar{\theta} = (\theta^{n+1} + \theta^n) / 2$. Then, we compute $(\nabla \cdot \mathbf{j})_\theta = \nabla \cdot \mathbf{j} - (\nabla \cdot \mathbf{j})_r$. Finally, we invert the nabla operators,
\begin{align}
{(\nabla \cdot \mathbf{j})_r}_{(i,j)} &= 3 \Bigg[ \frac{r_{i+1/2}^2 {j_r}_{(i+1/2,j)} - r_{i-1/2}^2 {j_r}_{(i-1/2,j)}}{r_{i+1/2}^3 - r_{i-1/2}^3} \Bigg] \ , \label{eq:div_jr_nabla} \\
{(\nabla \cdot \mathbf{j})_\theta}_{(i,j)} &= \frac{3}{2} \frac{r_{i+1/2}^2 - r_{i-1/2}^2}{r_{i+1/2}^3 - r_{i-1/2}^3} \times \nonumber \\
& \times \Bigg[ \frac{\sin \theta_{j+1/2} {j_\theta}_{(i,j+1/2)} - \sin \theta_{j-1/2} {j_\theta}_{(i,j-1/2)}}{\cos \theta_{j-1/2} - \cos \theta_{j+1/2}} \Bigg] \ , \label{eq:div_jt_nabla}
\end{align}
to find the current components. The inversion of ${(\nabla \cdot \mathbf{j})_\theta}_{(i,j)}$ is simple, because the second term in the square brackets of Eq.~\eqref{eq:div_jt_nabla} is always zero given that the particle motion is restricted to cell $(i,j)$. The same is applicable to the inversion of ${(\nabla \cdot \mathbf{j})_r}_{(i,j)}$ for most particle positions in cell $(i,j)$; however, due to the fact that the particle expands with $r_p$, it can deposit current at the grid position $(i-1/2,j)$ when $r_p$ is close to $r_{i}$. When this happens, we determine ${(\nabla \cdot \mathbf{j})_r}_{(i-1,j)}$ using Eq.~\eqref{eq:div_jr}, invert the corresponding operator to obtain ${j_r}_{(i-1/2,j)}$ and use it to solve for ${j_r}_{(i+1/2,j)}$ in Eq.~\eqref{eq:div_jr_nabla}. \textcolor{coralpink}{When particles cross two cells from $\mathbf{x}^n$ to $\mathbf{x}^{n+1}$, we split} their trajectory such that each split is within a single cell, and apply the method described \textcolor{coralpink}{before} to each trajectory split. \textcolor{coralpink}{The same strategy is applied in the algorithms proposed in~\citet{villasenor_1992} and \citet{esirkepov_2001}.} This method does not impose any restriction on the azimuthal current component, which we take to be simply ${j_\phi}_{(i,j)} = \rho_{(i,j)} v_\phi$, where $v_\phi$ is the macro-particle velocity in the azimuthal direction.

\textcolor{coralpink}{Finally, we note that Eqs.~\eqref{eq:shape_dep} and \eqref{eq:div_jr_nabla}-\eqref{eq:div_jt_nabla} can also be derived by applying the algorithms in~\citet{villasenor_1992} or \citet{esirkepov_2001} (in first-order) in a Cartesian logical space with the spherical coordinates metric. However, the special radial integration rule described in this section to account for particle shrinking/expansion should be included to ensure that those algorithms conserve charge to machine precision.}

\begin{figure*}[t]
\centering
\includegraphics[width=4.1in]{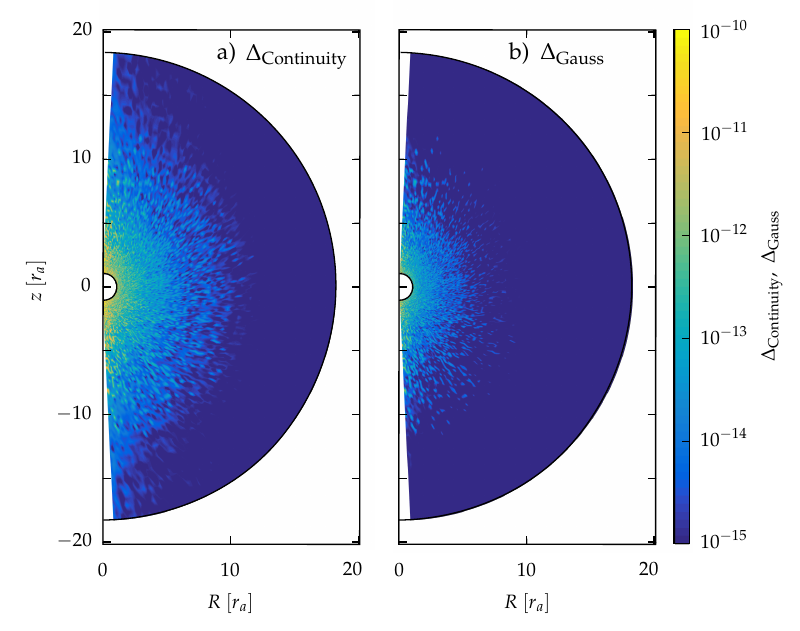}
\caption{Current deposition benchmarks, showing that both a) the continuity equation and b) Gauss' law are satisfied to machine precision.}
\label{fig:spherical_cont_gauss}
\end{figure*}

We have benchmarked the current deposition method presented here by initializing particles all over the simulation domain with a random velocity, depositing their current over a time step $\Delta t$ and evaluating
\begin{align}
\Delta_\mathrm{Continuity} & = \frac{\Delta t}{\rho_{(i,j)}} \left( \frac{\partial \rho_{(i,j)}}{\partial t} + (\nabla \cdot \mathbf{j})_{(i,j)} \right) \ , \\
\Delta_\mathrm{Gauss} & = \frac{1}{\rho_{(i,j)}} \left( (\nabla \cdot \mathbf{E})_{(i,j)} - 4 \pi \rho_{(i,j)} \right) \ .
\end{align}
Both $\Delta_\mathrm{Continuity}$ and $\Delta_\mathrm{Gauss}$ should be zero if the continuity equation and Gauss' law are satisfied. Figure~\ref{fig:spherical_cont_gauss} shows that these quantities are both of the order of $10^{-15} - 10^{-11}$, \textit{i.e.}, of the order of machine precision. The value of both $\Delta_\mathrm{Continuity}$ and $\Delta_\mathrm{Gauss}$ tends to be larger closer to the star, due to the larger number of operations subject to round-off errors in this region, caused by particles crossing more cell boundaries and depositing their current in more than one cell. We have verified that the accuracy of the method is maintained over multiple time steps by ensuring that the evolution of the grid integrals of $\Delta_\mathrm{Continuity}$ and $\Delta_\mathrm{Gauss}$ remain at machine precision level.

\textcolor{coralpink}{This current deposition method thus accurately conserves charge, avoiding the need for other correcting algorithms. It is also inexpensive, since most factors in Eqs.~\ref{eq:div_jtot}-\ref{eq:div_jt_nabla} can be precomputed and reused throughout a simulation.}

\subsection{Typical scales and normalizations}
\label{sec:spherical_normalizations}

In the benchmarks presented above, the normalization units of distances, times, and fields varied according to what best suits the \textcolor{coralpink}{respective tests}. However, for pulsar magnetosphere simulations, we adopt a common normalization that we introduce here. We choose to normalize distances to the stellar radius $r_*$ and times to $r_*/c$. Electric and magnetic fields are normalized to $m_e c^2 / e r_*$, however we typically represent them in units of $e n_\mathrm{GJ} r_*$, where $n_\mathrm{GJ} = \Omega B_* / 2 \pi e c$ is the surface Goldreich-Julian (GJ)~\citep{goldreich_julian_1969} particle number density. The GJ density also defines a typical frequency $\omega_{p, \mathrm{GJ}} = \sqrt{4 \pi e^2 n_\mathrm{GJ} / m_e}$ and an electron skin depth $d_{e, \mathrm{GJ}} = c / \omega_{p, \mathrm{GJ}}$. The time step and grid spacing are chosen to resolve these temporal and spatial scales, respectively.

In pulsar magnetosphere simulations, the main parameter responsible for setting the typical temporal, spatial and energy scales is the normalized value of the surface magnetic field, $B_* (e r_* / m_e c^2)$. For realistic parameters, $B_* \simeq 10^{12}$~G and $r_* \simeq 10$~km, we have $B_* (e r_* / m_e c^2) \sim 10^{15}$. Global simulations are not feasible with such values, since they would have to resolve scales of the order of $\sim$ tens of $r_*$ down to $d_{e, \mathrm{GJ}} \sim 10^{-7}~r_*$. For this reason, we use more modest values of $B_* (e r_* / m_e c^2) \sim 10^3 - 10^6$, such that we respect the ordering \textcolor{coralpink}{in these objects,} $\Omega \ll \omega_{p, \mathrm{GJ}} \ll \omega_c$, where $\omega_c = e B_* / m_e c$ is the cyclotron frequency associated with a field magnitude $B_*$.

\section{Global simulations of pulsar magnetospheres}
\label{sec:global}

In this Section, we present global PIC simulations of pulsar magnetospheres \textcolor{coralpink}{obtained with the OSIRIS framework~\citep{fonseca_2002, fonseca_2008}}. We start by allowing \textcolor{coralpink}{electron-positron} pairs to be artificially and abundantly injected in our simulations, and then make increasingly realistic assumptions about the plasma supply processes, in particular regarding the regions of space where pair cascades operate, and the separation between kinetic and system scales. 

All simulations presented here have a similar initial configuration: the system starts in vacuum and with an initial dipolar magnetic field of polar surface magnitude $B_*$, \textit{i.e.}, $B_r (r, \theta) = B_* (r_* / r)^3 \cos \theta$ and $B_\theta (r, \theta) = (1/2) B_* (r_* / r)^3 \sin \theta$. The inner radial boundary is treated as a rotating conductor of angular velocity $\boldsymbol{\Omega} = \Omega \mathbf{\hat{z}}$; at the surface of the neutron star, we impose the co-rotation electric field $\mathbf{E} = - (\mathbf{v}_\mathrm{rot} \times \mathbf{B}) / c$, with $\mathbf{v}_\mathrm{rot} = \boldsymbol{\Omega} \times (r_* \mathbf{\hat{r}})$. In all simulations, we consider the stellar rotation frequency to be initially zero and increase it linearly over a time $t_\mathrm{rise} c / r_* = 1.5$ to $\Omega r_* / c = 0.125$. For times $t > t_\mathrm{rise}$, the stellar frequency is kept constant. The stellar period is $T = 2\pi / \Omega = 50~r_*/c$ and the light-cylinder radius is $R_\mathrm{LC} / r_* = 8$. All simulations use also $r_\mathrm{min} / r_* = 1$ and $r_\mathrm{max} / r_* = 20$, such that the plasma dynamics can be captured up to $r / R_\mathrm{LC} > 2$. The value of $B_*$ is chosen to satisfy the ordering $\Omega \ll \omega_{p, \mathrm{GJ}} \ll \omega_c$ described in Sect.~\ref{sec:spherical_normalizations} while maintaining simulations numerically feasible. This choice and others regarding e.g., grid resolution vary according to the injection scheme and parameter regime under study, and are detailed alongside the corresponding simulations.

\subsection{Volume injection}
\label{sec:global_vol}

In this section, we inject plasma everywhere in the simulation domain where the local electric field component parallel to the magnetic field satisfies the condition $E_\parallel c / r_* \Omega B_* > k_\mathrm{lim}$, where $k_\mathrm{lim}$ is a constant. Similar injection criteria have been used in \citet{belyaev_2015}, whereas in \citet{philippov_2014, kalapotharakos_2018, brambilla_2018} plasma is only injected if the local magnetization is also above a given threshold. Physically, this injection scheme is equivalent to assuming that \textcolor{coralpink}{electron-positron} pair cascades may develop wherever $E_\parallel$ is sufficiently large, \textit{i.e.}, it neglects any role of the local magnetic field magnitude or curvature. Since all fields (and in particular $E_\parallel$) decay with $r$, the choice of $k_\mathrm{lim}$ can also be interpreted as a spatial limitation to the plasma supply: infinitely small values of $k_\mathrm{lim}$ allow plasma to be injected up to $r \gg r_*$, whereas $k_\mathrm{lim} \sim 1$ restricts the plasma supply to radii $r \sim r_*$. A macro-electron-positron pair carrying a number density $n_\mathrm{vol} = k_\mathrm{vol} E_\parallel / e r_*$, with $k_\mathrm{vol} = 0.2$, is injected at rest in each cell and time step in which the injection condition is met. The choice of $k_\mathrm{vol}$ is such that a few macro-particles are required to supply the charge density that screens $E_\parallel$ and stops the injection. We can also interpret $k_\mathrm{vol}$ as a parameter proportional to the local GJ density, since $E_\parallel / e r_* \sim n_\mathrm{GJ}$. In all the simulations presented in this section, $B_* e r_* / m_e c^2 = 8 \times 10^3$, $N_r \times N_\theta = 1000^2$ and $\Delta t c / r_* = 10^{-3}$. In these conditions, $c / \omega_{p, \mathrm{GJ}} r_* \simeq 0.022$, whereas the minimum grid spacing is $\min (\Delta r_i) / r_* \simeq 0.003$.

\begin{figure*}[t]
\centering
\includegraphics[width=4.8in]{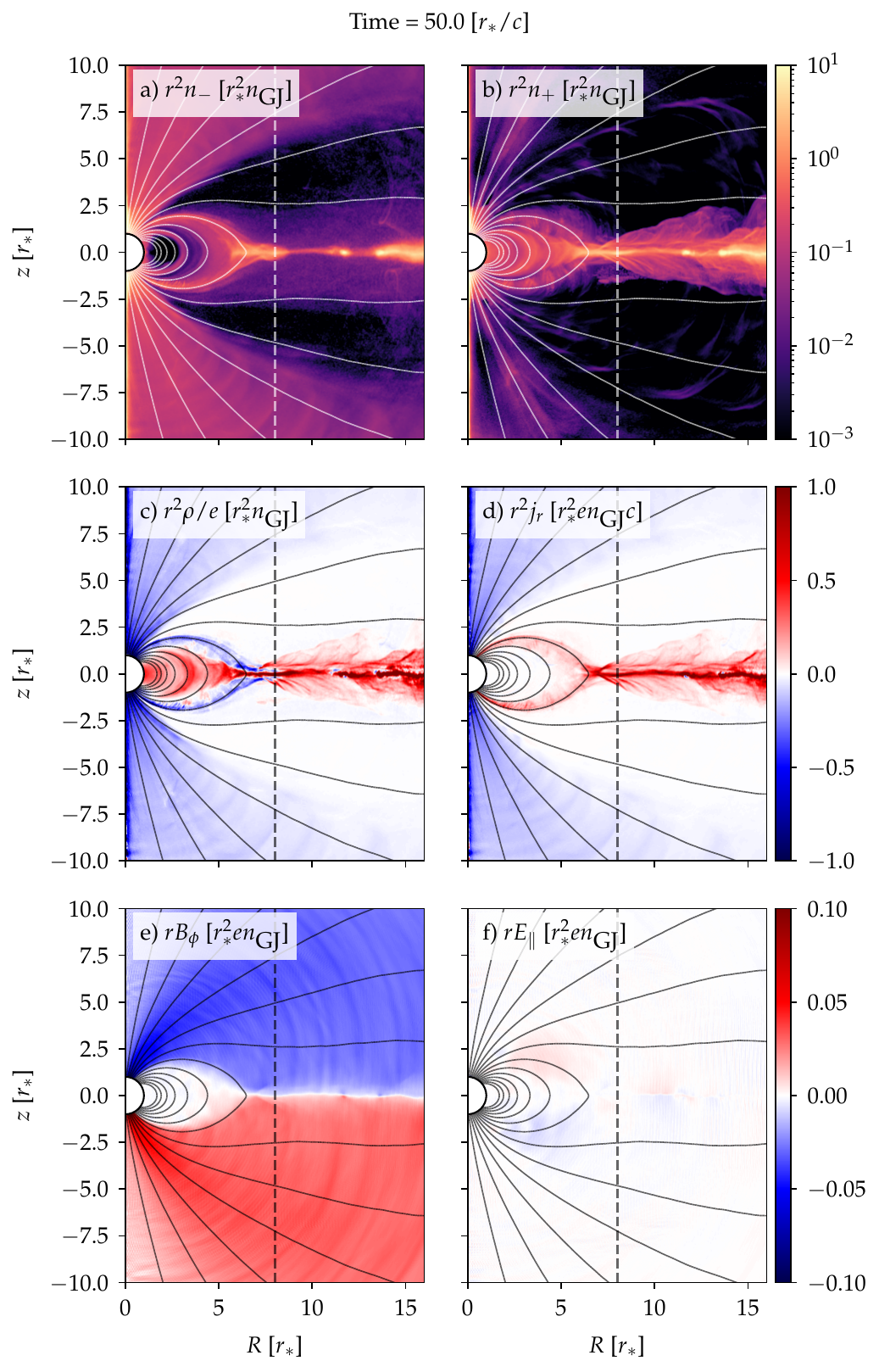}
\caption{Force-free magnetosphere obtained with volume injection. Panels a-f show the electron and positron density, total charge density, radial current density, azimuthal magnetic field and electric field component parallel to the local magnetic field, respectively. Quantities are multiplied by powers of $r$ to enhance large radii features. White/black solid lines represent magnetic field lines, and vertical dashed lines show the location of the light-cylinder.}
\label{fig:vol-ff-overview}
\end{figure*}

In Fig.~\ref{fig:vol-ff-overview}, we present an overview of the quasi-steady-state solution obtained with $k_\mathrm{lim} = 0.005$. This solution is achieved after a time $\sim 25~r_*/c \sim T / 2$\footnote{This is not a universal result. In fact, the time required by the system to achieve a steady-state (or quasi steady-state) solution varies with the injection scheme, the stellar ramp-up time $t_\mathrm{rise}$ and other initial and/or boundary conditions.}. In the first half stellar period, the simulation undergoes a transient stage in which the vacuum co-rotation fields are established and plasma is created. The solution presented in Fig.~\ref{fig:vol-ff-overview} resembles the canonical force-free regime of pulsar magnetospheres: the magnetosphere is divided in two regions permeated by closed and open magnetic field lines (shown in white/black solid lines in all panels), with the last closed field line crossing the equatorial plane at the light-cylinder radius (shown in a white/black dashed vertical line in all panels). The open and closed field line regions are respectively negatively and positively charged, even if electrons and positrons exist in both regions --- see Fig.~\ref{fig:vol-ff-overview}a-c, showing the electron and positron number density and the total charge density, respectively. As shown in Fig.~\ref{fig:vol-ff-overview}d, a negative radial current density $j_r$ (blue) is conducted from the polar regions and along the open field lines, which is compensated by return current layers (red) established on the last closed field line. The return current layers are connected with each other at a distance $r \simeq R_\mathrm{LC}$ on the equatorial plane, where the poloidal magnetic field lines resemble a Y shape. A radial current density layer extends along the equatorial plane to large distances, supporting a strong gradient in the toroidal magnetic field component $B_\phi$, illustrated in Fig.~\ref{fig:vol-ff-overview}e. The poloidal magnetic field lines have also opposite polarity in opposite sides of this equatorial current layer, and reconnect sporadically, leading to the formation of outflowing plasmoids --- see the large density structures at $r / r_* \simeq 12$ in Fig.~\ref{fig:vol-ff-overview}a-b. The plasma supply in this simulation is large enough such that $E_\parallel$ is effectively screened in the whole simulation domain, as shown in Fig.~\ref{fig:vol-ff-overview}f, and thus lies well within the assumptions of the force-free regime for pulsar magnetospheres.

The quasi-steady-state shown in Fig.~\ref{fig:vol-ff-overview} is sustained via intermittent injection, mainly along the return current layers. In these regions, $E_\parallel$ is less efficiently screened, leading to the injection of plasma which, in turn, screens the field as it flows along the return current layers. As we shall demonstrate, this intermittency has a period of $\simeq 0.3-0.5~T$, and it may play a significant role in the temporal evolution of the magnetospheric state. However, for $k_\mathrm{lim} = 0.005$ the solution never deviates significantly from the force-free regime.

\begin{figure*}[t]
\centering
\includegraphics[width=5.66in]{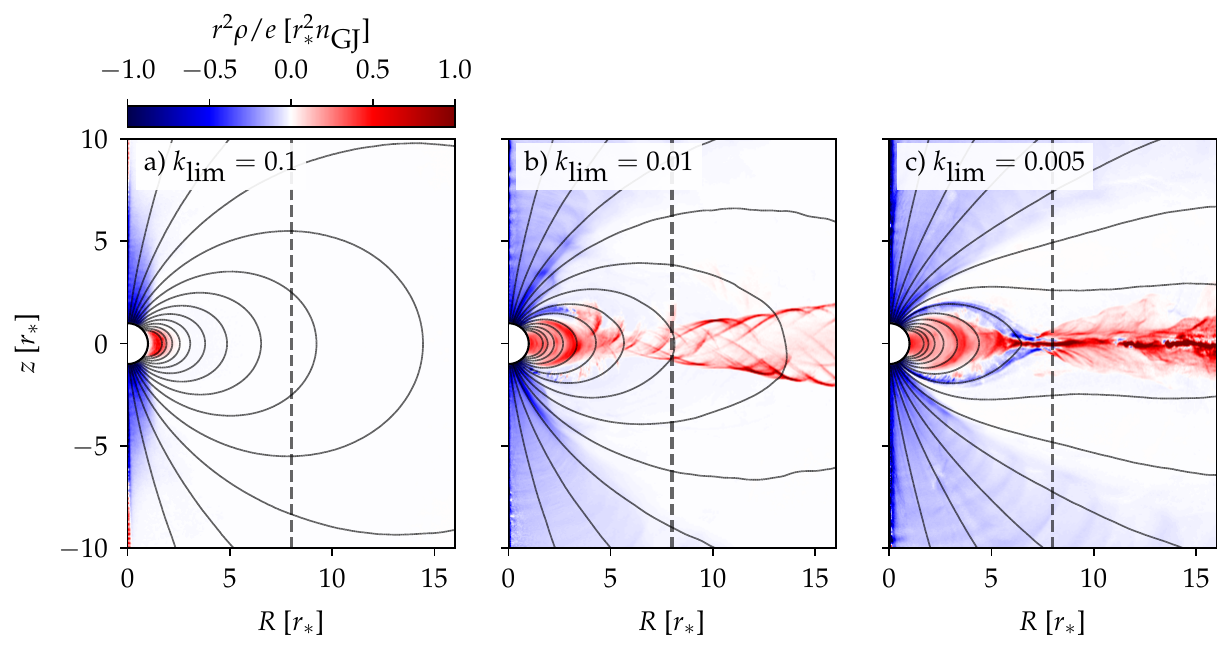}
\caption{Magnetospheric solutions obtained with volume injection. The panels show the total charge density after a stellar rotation period.}
\label{fig:vol-rho-comp}
\end{figure*}

\textcolor{coralpink}{In order to demonstrate how the magnetospheric solution changes with $k_\mathrm{lim}$}, in Fig.~\ref{fig:vol-rho-comp} we compare the total charge density of the solutions obtained with $k_\mathrm{lim} = \{0.005, 0.01, 0.1\}$. We recall that $k_\mathrm{lim}$ is the minimum value of $E_\parallel c / r_* \Omega B_*$ for which we inject plasma. It is clear that the force-free regime is only observed for $k_\mathrm{lim} = 0.005$. For $k_\mathrm{lim} = 0.01$, the equatorial current sheet (positively charged region at $r \gtrsim R_\mathrm{LC}$) is wide and the return current layers are not positively charged everywhere, and for $k_\mathrm{lim} = 0.1$ the solution does not even produce an outflow. In fact, by increasing $k_\mathrm{lim}$, we are limiting the plasma supply to regions closer and closer to the stellar surface. This can be understood by noting that this parameter compares the local $E_\parallel$ with the reference value $\Omega B_* r_* / c$ (\textit{i.e.}, the surface magnitude of the electric field in vacuum). Since the typical magnitude of $E_\parallel$ decreases with $r$, decreasing $k_\mathrm{lim}$ limits plasma injection to smaller radii. In the $k_\mathrm{lim} = 0.01$ run, this supply occurs only up to radii $r / r_* \simeq 3$, and the solution shows the same intermittency observed for $k_\mathrm{lim} = 0.005$. However, the injection stage is not as efficient in this case, and the equatorial outflow is not dense enough to produce a thin current sheet. For $k_\mathrm{lim} = 0.1$, only regions close to the surface can initially fulfil the injection criteria, and no plasma is supplied to large radii. The system relaxes in this case to a fully charge-separated configuration, with only electrons (positrons) in the poles (equatorial region). This solution is often denominated as the disk-dome or electrosphere solution~\citep{jackson_1976, krause-polstorff_1985}. In the charged regions, the electric field is screened, injection ceases and no plasma outflows are formed.

\begin{figure*}[t]
\centering
\includegraphics[width=4in]{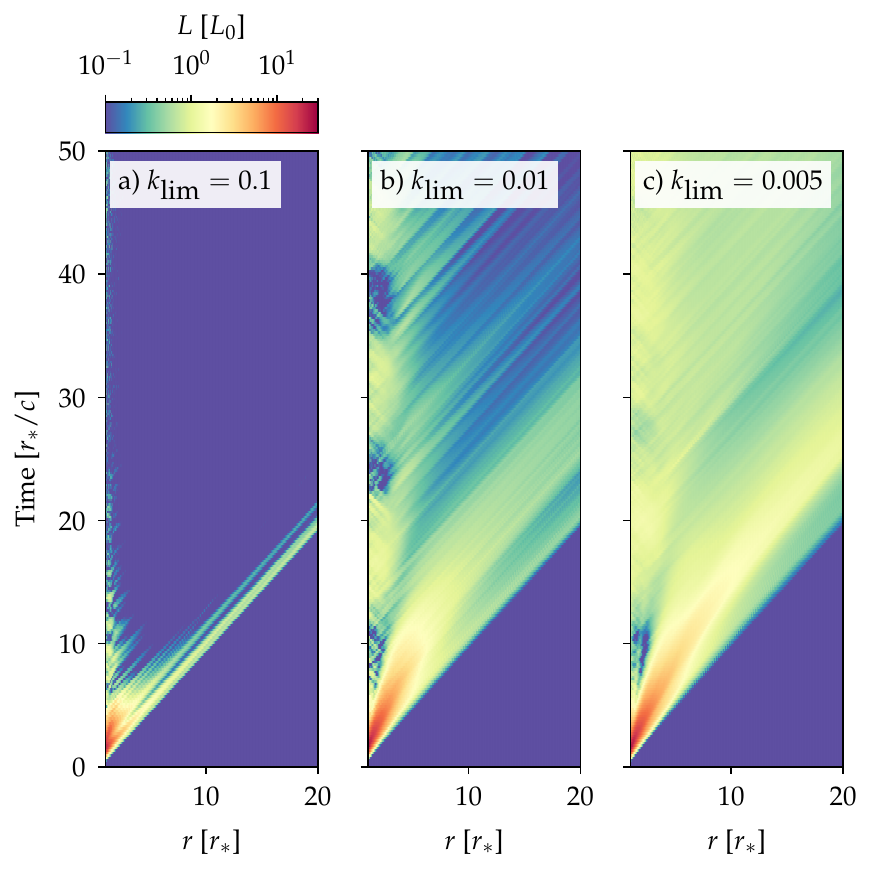}
\caption{Poynting flux in simulations with volume injection. Values are normalized to the theoretical value $L_0 = \mu^2 \Omega^4 / c^3$.}
\label{fig:vol-L-comp}
\end{figure*}

\begin{figure*}[t]
\centering
\includegraphics[width=4in]{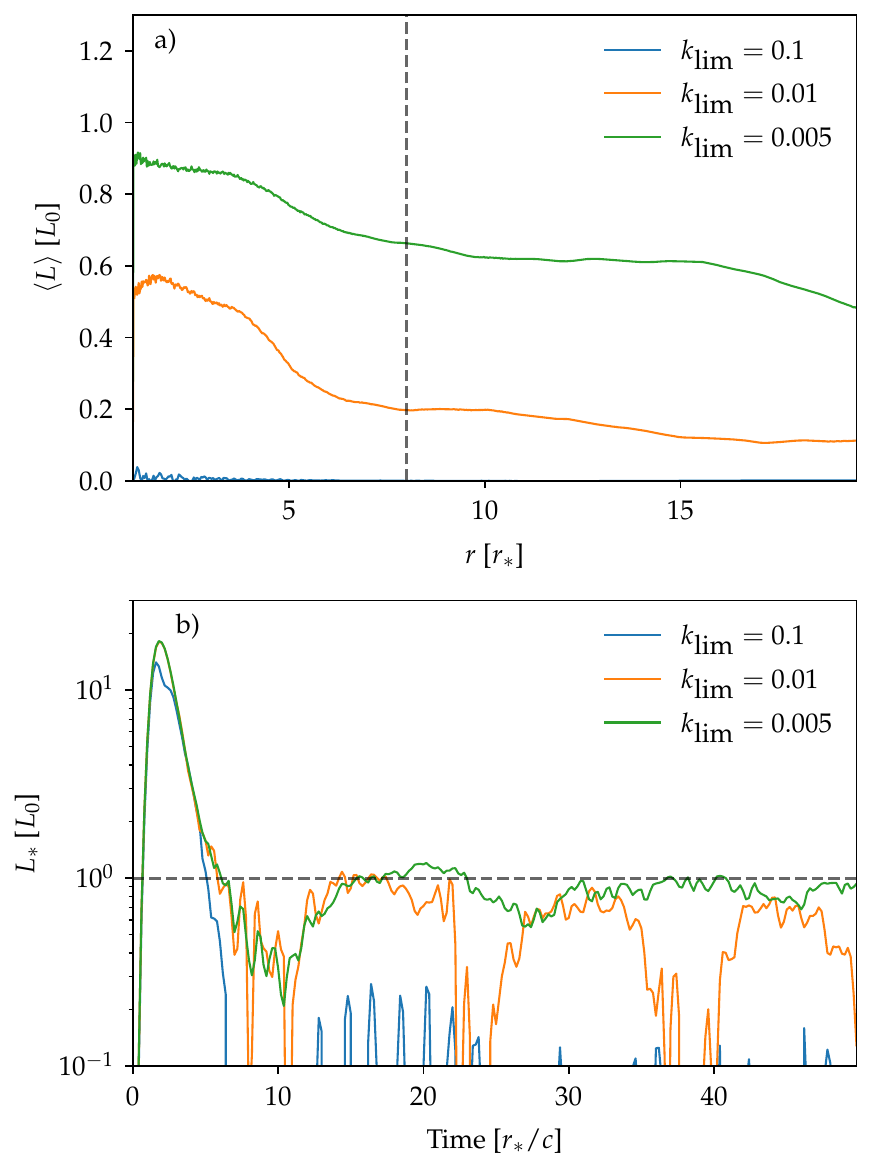}
\caption{Radial and temporal dependencies of Poynting flux in simulations with volume injection. a) shows the time-averaged luminosity $\langle L \rangle$ as a function of $r$ after a stellar rotation period, and b) shows the temporal evolution of the surface Poynting flux $L_*$. The dashed lines in a) and b) identify the light-cylinder radius and the theoretical surface Poynting flux $L_0 = \mu^2 \Omega^4 / c^3$, respectively.}
\label{fig:vol-L-comp-avgs}
\end{figure*}

\textcolor{coralpink}{An important property of the magnetospheric solution is} the integrated Poynting flux $L(r)$, defined as
\begin{equation}
L(r) = \frac{c}{2} \int_0^\pi (\mathbf{E} \times \mathbf{B})_r \ r^2 \sin \theta \mathrm{d}\theta  \ .
\end{equation}
Figure~\ref{fig:vol-L-comp} shows $L(r)$ as a function of time for the three simulations described before. This quantity is normalized to the theoretical value of the spindown luminosity, $L_0 = \mu^2 \Omega^4 / c^3$, with $\mu = B_* r_*^3$. We observe a large spindown at early times for all simulations, which is a consequence of the initial transient stage. After this transient, the $k_\mathrm{lim} = 0.1$ simulation converges to a surface Poynting flux $L_* / L_0 \ll 1$, which is a consequence of the inactivity of disk-dome solution. On the contrary, the simulations with lower $k_\mathrm{lim}$ have $L_* / L_0 \sim 1$. The Poynting flux remains approximately constant within the light-cylinder for these runs, and decays with $r$ for $r > R_\mathrm{LC}$, which is a signature of the conversion from magnetic to kinetic energy due to magnetic reconnection in the equatorial plane. The surface Poynting flux shows variations of periodicity $0.3 - 0.5~T$, which are correlated with the intermittency of the solution identified above in this section. The time-averaged radial dependence of the luminosity $\langle L \rangle$ after a stellar period and the temporal dependence of $L_*$ is shown in Fig.~\ref{fig:vol-L-comp-avgs}.

The simulations presented in this section show that the efficiency of the plasma supply critically determines the global structure of the pulsar magnetosphere. It is expected that pulsar magnetoshere is in a regime close to the force-free configuration identified with $k_\mathrm{lim} = 0.005$ or lower. However, pair production cannot operate in all regions of the magnetosphere, in particular at radii comparable to the light-cylinder radius. It is then important to assess if more realistic injection and/or pair production schemes can provide the plasma supply required for the magnetosphere to be in the force-free regime. In the next sections, we address this question by considering plasma supply schemes limited to regions close to the stellar surface.

\subsection{Surface injection}
\label{sec:global_sur}

In this section, we limit injection to occur only at the stellar surface. In doing so, we phenomenologically introduce the important role of the magnetic field amplitude in our treatment of the magnetospheric plasma supply. As in Sect.~\ref{sec:global_vol}, we do not allow particles to emit photons and/or pairs. We adopt two different criteria for the injection and vary the density and velocity of the surface-injected plasma. The parametrization of the plasma flow injected from the stellar surface is similar to that presented in~\citet{cerutti_2015}. However, our criteria for injection differ slightly from that work, that also assumes a minimum threshold for the local plasma magnetization. In all simulations presented in this section, we use $B_* e r_* / m_e c^2 = 8 \times 10^3$, $N_r \times N_\theta = 500^2$ and $\Delta t c / r_* = 3 \times 10^{-3}$.

\begin{figure*}[t]
\centering
\includegraphics[width=5.05in]{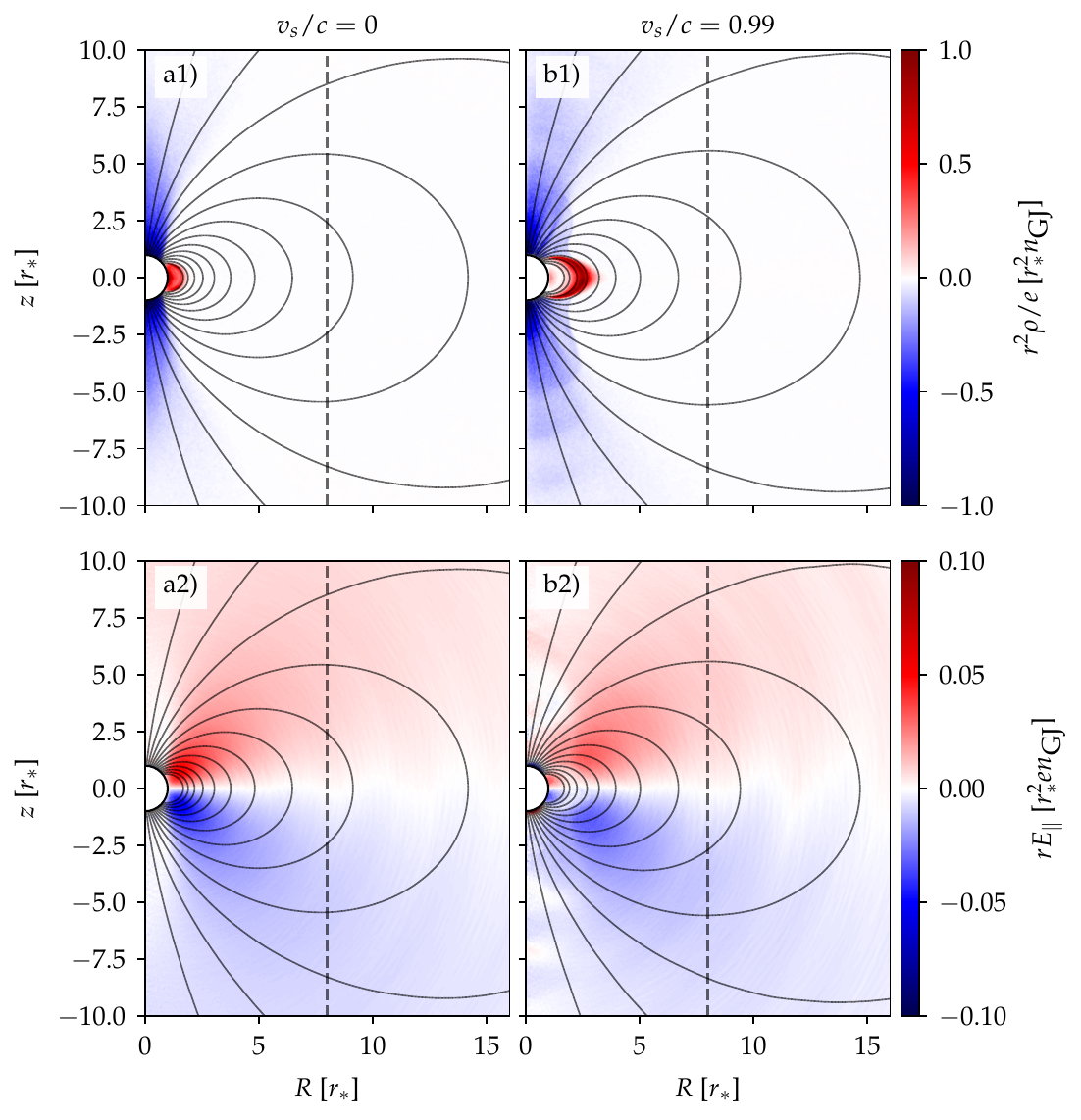}
\caption{Magnetospheric solutions obtained with surface injection proportional to $E_\parallel$. a1-2) show the total charge density and $E_\parallel$, respectively for a simulation with $v_\mathrm{s} / c = 0$ and b1-2) show the same for a simulation with $v_\mathrm{s} / c = 0.99$. Solid lines represent magnetic field lines, and vertical dashed lines show the location of the light-cylinder.}
\label{fig:sur-EdotB-comp}
\end{figure*}

The first injection criterion is based on the local value of $E_\parallel$. We inject a macro-electron-positron pair in each cell just above the stellar surface ($r = r_*$) that satisfies $E_\parallel c / r_* \Omega B_* > k_\mathrm{lim}$. In this case, we consider a fixed $k_\mathrm{lim} = 0.002$ and vary the properties of the injected pairs, namely their density $n_\mathrm{s} = k_\mathrm{s} n_\mathrm{GJ}$ and poloidal velocity $v_\mathrm{s}$. These pairs are also injected with a toroidal velocity that matches the local linear velocity of the stellar surface, $v_\phi = \Omega r \sin \theta$.

Despite the large range of injection parameters considered, $k_\mathrm{s} = n_\mathrm{s} / n_\mathrm{GJ} = \{0.2, 0.5, 1\}$ and $v_\mathrm{s} / c = \{0, 0.1, 0.5, 0.99 \}$, the solutions obtained for long times, $t / T \gtrsim 2$, always converge to the disk-dome solution identified in Sect.~\ref{sec:global_vol}. Figure~\ref{fig:sur-EdotB-comp} shows the charge density $\rho$ and $E_\parallel$ of two runs with $k_\mathrm{s} = n_\mathrm{s} / n_\mathrm{GJ} = 1$ and $v_\mathrm{s} = \{0, 0.99\}$ after a time $t / T \simeq 4$. After an initial transient, the system settles to a charge-separated solution and effectively screens $E_\parallel$ at the stellar surface, precluding further injection.

The second injection criterion does not depend on the local surface field conditions. Instead, injection is allowed in all cells above the stellar surface in which the combined local number density of positrons and electrons satisfies $n_+ + n_- < 5~n_\mathrm{GJ}$\textcolor{coralpink}{, to ensure that enough plasma exists everywhere to screen the local electric field parallel to the magnetic field}. We emphasize that $n_\mathrm{GJ} = \Omega B_* / 2 \pi e c$ is the pole GJ density and not its local value. This criterion allows injection to occur even if $E_\parallel \sim 0$, and is thus harder to motivate from first-principles arguments. Here, we shall interpret it as a means of \textcolor{coralpink}{producing a set plasma density over a layer near the stellar surface of width smaller than the local resolution of the simulation grid. In pulsars, such layer can be as small as $\sim 100$~m~\citep{ruderman_sutherland_1975}}. We consider that the injected electron-positron pairs carry a number density $n_\mathrm{s} = k_\mathrm{s} n_\mathrm{GJ}$ and poloidal velocity $v_\mathrm{s}$.

\begin{figure*}[t]
\centering
\includegraphics[width=4.05in]{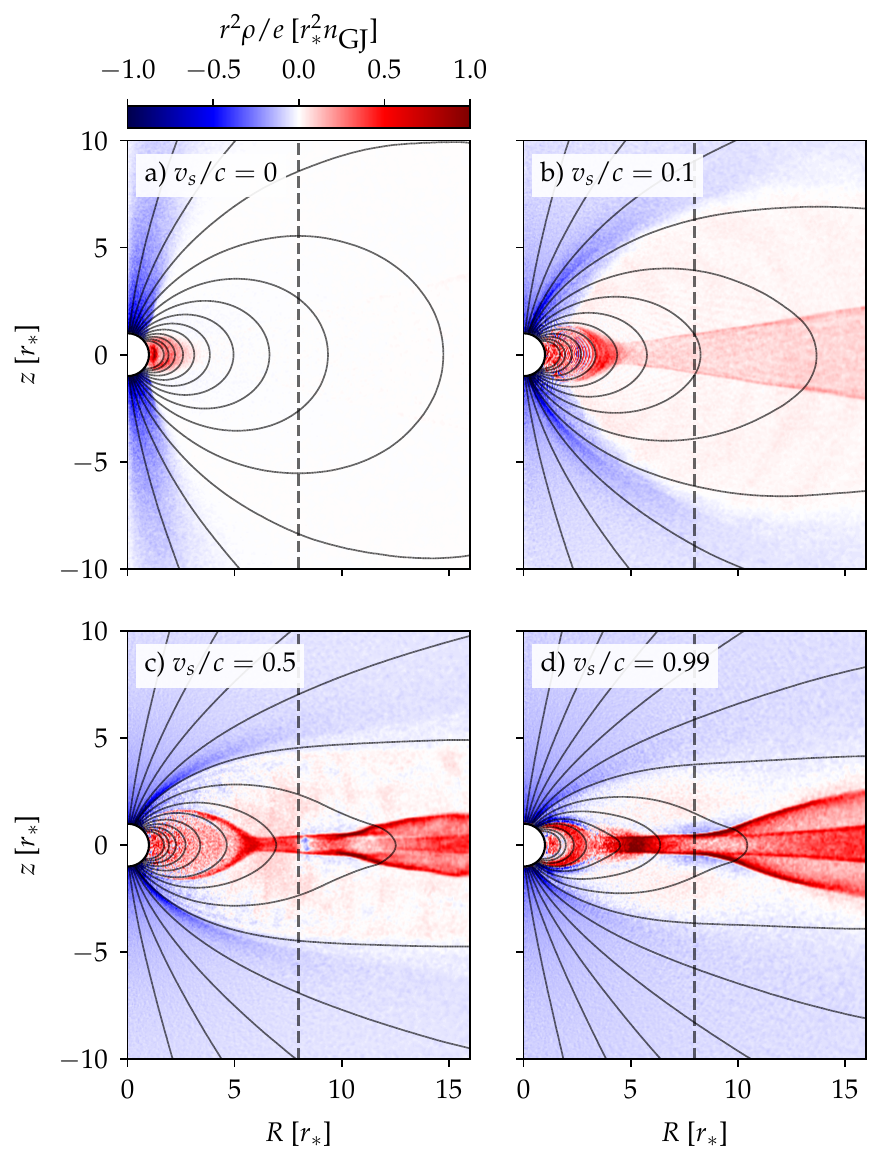}
\caption{Magnetospheric solutions obtained with surface injection proportional to $n_\mathrm{GJ}$ with fixed $k_\mathrm{s} = n_\mathrm{s} / n_\mathrm{GJ} = 0.2$ and varying $v_\mathrm{s} / c$.}
\label{fig:sur-nGJ-vs-rho-comp}
\end{figure*}

\begin{figure*}[t]
\centering
\includegraphics[width=4in]{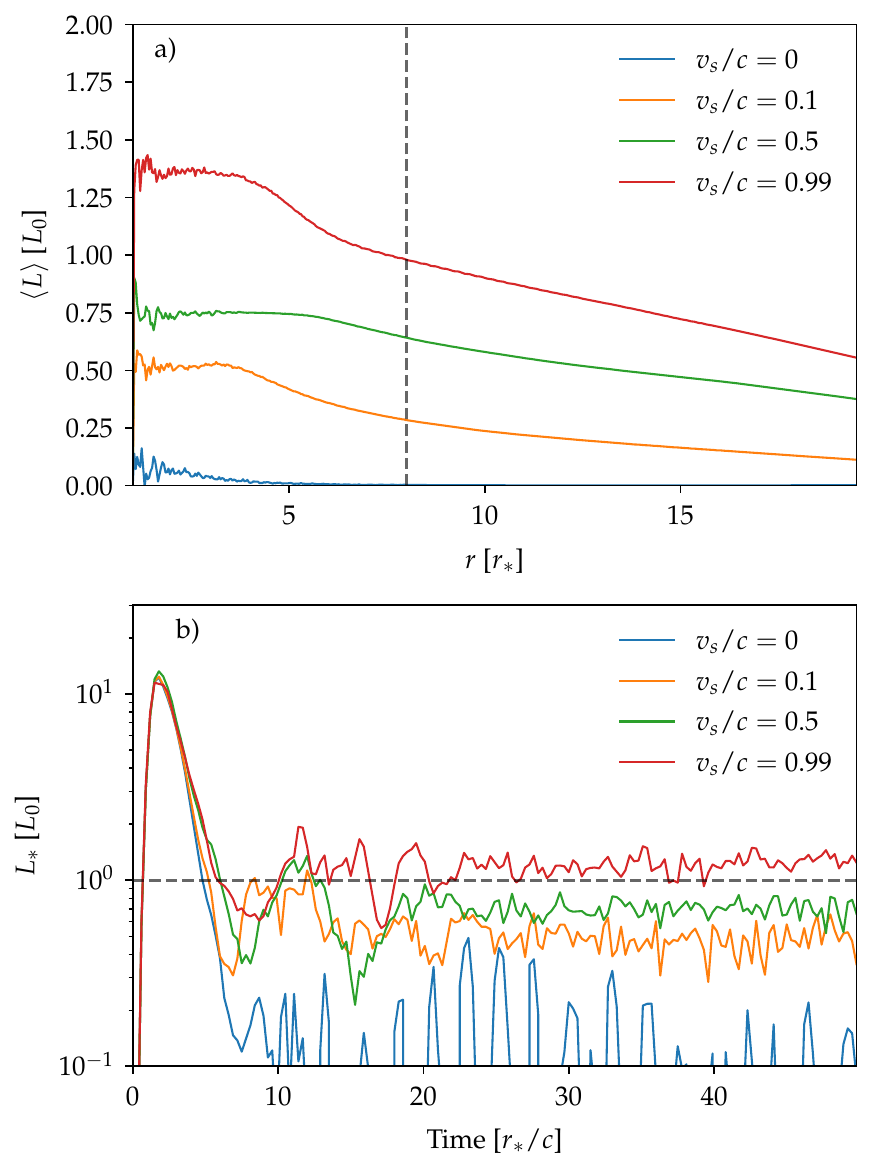}
\caption{Radial and temporal dependencies of Poynting flux in simulations with surface injection proportional to $n_\mathrm{GJ}$ with fixed $k_\mathrm{s} = n_\mathrm{s} / n_\mathrm{GJ} = 0.2$ and varying $v_\mathrm{s} / c$. a) shows the time-averaged luminosity $\langle L \rangle$ as a function of $r$ after a stellar rotation period, and b) shows the temporal evolution of the surface Poynting flux $L_*$. The dashed lines in a) and b) identify the light-cylinder radius and the theoretical surface Poynting flux $L_0 = \mu^2 \Omega^4 / c^3$, respectively.}
\label{fig:sur-nGJ-vs-L-avgs-comp}
\end{figure*}

In Fig.~\ref{fig:sur-nGJ-vs-rho-comp}, we show the charge density distribution of the solutions obtained for a fixed $k_\mathrm{s} = n_\mathrm{s} / n_\mathrm{GJ} = 0.2$ and varying $v_\mathrm{s}$ for a time $t / T = 1$. With $v_\mathrm{s} = 0$, the system converges to the electrosphere solution. \textcolor{coralpink}{Particles injected at early times develop a space-charge limited flow, driving $E_\parallel$ to zero near the stellar surface and thus inhibiting freshly injected particles to be pulled away from or towards the star.} For $v_\mathrm{s} > 0$, we observe that the system develops a positively charged outflow along the equatorial plane. This outflow occurs in a narrower current sheet for larger values of $v_\mathrm{s}$, which can be understood as a mechanism to support the stronger toroidal magnetic field driven by the stronger poloidal currents of these regimes. However, we do not observe a current sheet as thin as that characteristic of the force-free regime. Instead, the current sheet remains wide even for $v_\mathrm{s} / c = 0.99$. This may indicate that the plasma launched into this region is not dense enough, a question that we address below in this section.  

Figure~\ref{fig:sur-nGJ-vs-L-avgs-comp} shows the time-averaged Poynting flux produced by the simulations described above with surface injection as a function of the radial coordinate $r$ and its surface value as a function of time. We see once again that an electrosphere solution ($v_\mathrm{s} / c = 0)$ produces no spindown luminosity, and that it increases overall with increasing $v_\mathrm{s}$. The same decrease for $r > R_\mathrm{LC}$ observed in Sect.~\ref{sec:global_sur} is observed here. We note that the $v_\mathrm{s} / c = 0.99$ run shows a surface Poynting flux larger than $L_0$, which is a consequence of the smaller size of the co-rotation region (and thus a smaller effective light-cylinder radius and larger effective $L_0$).

\begin{figure*}[t]
\centering
\includegraphics[width=5.66in]{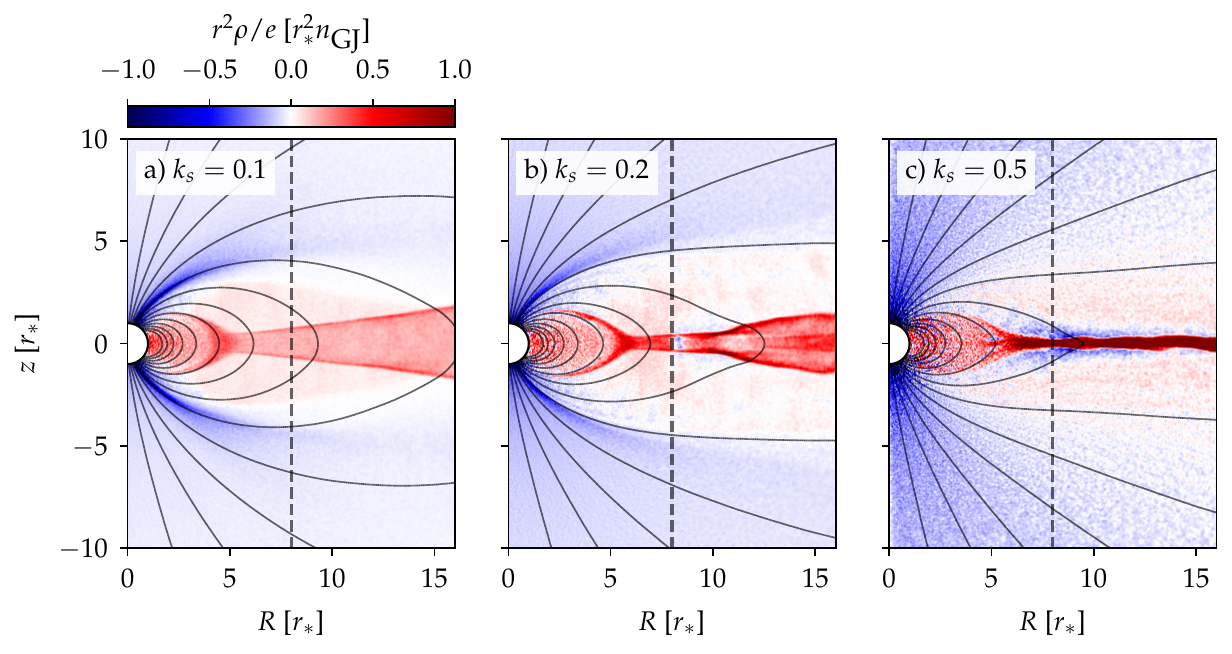}
\caption{Magnetospheric solutions obtained with surface injection proportional to $n_\mathrm{GJ}$ with fixed $v_\mathrm{s}$ and varying $k_\mathrm{s} = n_\mathrm{s} / n_\mathrm{GJ}$.}
\label{fig:sur-nGJ-ks-rho-comp}
\end{figure*}

We have also performed a set of simulations with fixed $v_\mathrm{s} / c = 0.5$ and varying $k_\mathrm{s} = n_\mathrm{s} / n_\mathrm{GJ} = \{0.1, 0.2, 0.5\}$. The charge density obtained in the steady-state (or quasi-steady-state) of these simulations is shown in Fig.~\ref{fig:sur-nGJ-ks-rho-comp}. These results confirm that the denser the injected plasma is, the more the solution approaches the force-free regime (see in particular the solution obtained for $k_\mathrm{s} = 0.5$). This injection density requirement seems to be critical in the launching of large density plasma to large radii, in particular along the return current layers, that connect the surface to the equatorial current sheet.

In summary, some of the parameters used in simulations presented in this section yield active magnetospheric solutions, with $L_* / L_0 \sim 1$ and a global configuration similar to the force-free regime. This is consistent with the results presented in~\citet{cerutti_2015}. However, it is hard to motivate the injection criteria and the choice of numerical parameters required to observe such regime.

\subsection{Pair production}
\label{sec:global_pp}

The results presented in Sects.~\ref{sec:global_vol} and \ref{sec:global_sur} are in good agreement with similar previous works. In particular, both \citet{philippov_2014} and \citet{cerutti_2015} observe a transition from electrosphere to active solutions with more abundant plasma supply. While in \citet{philippov_2014} pairs are injected up to large radii, in \citet{cerutti_2015} only surface injection is considered, showing trends with $k_\mathrm{s}$ and $v_\mathrm{s}$ very similar to our results.

The convergence to a force-free regime in the asymptotic limit of large plasma supply with both volume and surface injection is reassuring. However, an important question remains open when translating global simulations with volume and surface injection schemes to realistic systems: how is this plasma supplied, if strong field pair production operates efficiently only near the stellar surface? Is this pair production channel enough to supply the plasma to fill the whole magnetosphere?

In young and rapidly rotating pulsars (e.g., the Crab pulsar and other gamma-ray pulsars), pairs can also be created via the $\gamma$-$\gamma$ channel. In this process, for which the cross-section peaks at around a center of mass energy $\sim 2~m_e c^2$, gamma-rays produced via synchrotron emission and/or inverse Compton scattering in the equatorial current sheet collide with photons from a low energy bath, producing pairs. However, slower pulsars are not expected to have a sufficiently dense low-energy photon bath for this process to be relevant, and strong field pair production remains the main plasma supply channel.

In this section, we use global simulations that include pair production only near the stellar surface to understand whether it can provide enough plasma to maintain an active magnetospheric solution. We use the heuristic pair production model described in~\citet{cruz_2021, cruz_2022}, in which a lepton emits a pair of combined energy $\gamma_\mathrm{pair} m_e c^2$ whenever it achieves a threshold Lorentz factor $\gamma_\mathrm{thr}$. We keep the ratio $\gamma_\mathrm{thr} / \gamma_\mathrm{pair}$ constant, and vary the ratio $\eta \equiv \gamma_\mathrm{max} / \gamma_\mathrm{thr}$, where $\gamma_\mathrm{max} = e \Phi_\mathrm{pc} / m_e c^2$ is the maximum energy achievable by the particles in the voltage $\Phi_\mathrm{pc} = B_* r_*^3 \Omega^2 / c^2$ induced by the rotating star across the polar cap. In general, $\gamma_\mathrm{pair} \ll \gamma_\mathrm{thr} \ll \gamma_\mathrm{max}$ in real systems; however, it is very hard to achieve a large separation between these scales in global PIC simulations. For instance, previous works, considering a similar pair production model~\citep{chen_2017, philippov_2015b}, have used $\eta \sim 10$ and $\gamma_\mathrm{thr} / \gamma_\mathrm{pair} \sim 2$, which severely limits the efficiency of the pair cascades and the plasma multiplicity. In this Section, we present simulations with fixed $\gamma_\mathrm{pair} = 16$ and $\gamma_\mathrm{thr} = 25$ and a range of large values of $\eta$. We achieve this by controlling the surface magnetic field amplitude $B_*$. In doing this, besides increasing the scale separation between pair production and the dynamical scales, we also decrease the plasma kinetic scales. For this reason, we adopt a varying number of grid cells and time steps in our simulations to be able to resolve these scales. For $\eta = 5$ we use $N_r \times N_\theta = 500^2$ and $\Delta t c / r_* = 3 \times 10^{-3}$, for $\eta = \{ 25, 50 \}$ we use $N_r \times N_\theta = 1000^2$ and $\Delta t c / r_* = 10^{-3}$ and for $\eta = \{ 100, 150 \}$ we use $N_r \times N_\theta = 2000^2$ and $\Delta t c / r_* = 5 \times 10^{-4}$.

\begin{figure*}[t]
\centering
\includegraphics[width=4.05in]{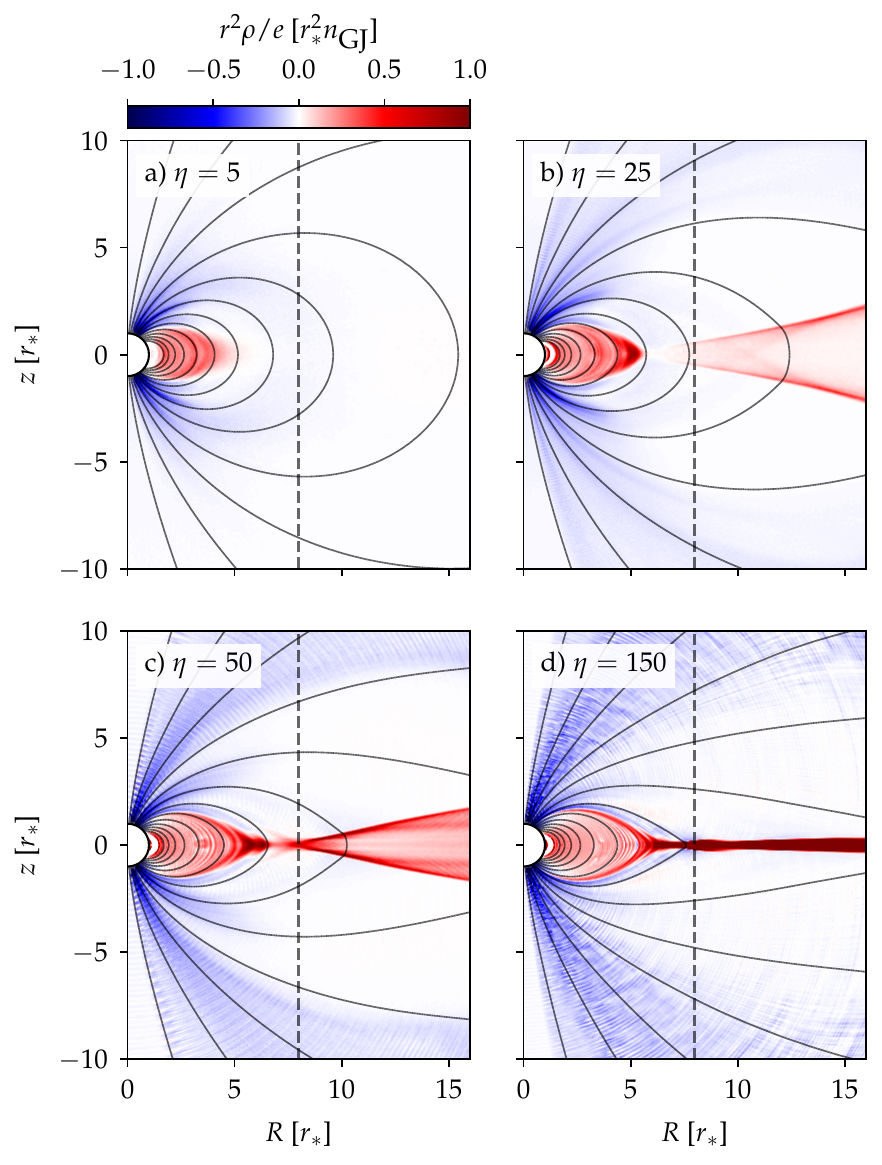}
\caption[Magnetospheric solutions obtained with pair production.]{\label{fig:pp-rho-comp}Magnetospheric solutions obtained with pair production. Panels a-d) show the total charge density for simulations with $\eta = \{5, 25, 50, 150\}$. Solid lines represent magnetic field lines, and vertical dashed lines show the location of the light-cylinder.}
\end{figure*}

In order to mimic the relevance of the large magnetic field required for pair production to occur, we limit pair production to only occur at radii $r / r_* < 3$. We also forbid pair production for $\theta < 0.01$, \textcolor{coralpink}{to reproduce the suppression of the corresponding QED cross-section in this region~\citep{cruz_2021c}}. Seed electron-positron pairs are provided at the stellar surface whenever $E_\parallel c / r_* \Omega B_* > k_\mathrm{lim}$, with $k_\mathrm{lim} = 0.1$. Each pair is injected at rest and carrying a density $n_\mathrm{s} = k_\mathrm{s} E_\parallel / e r_*$, with $k_\mathrm{s} = n_\mathrm{s} / n_\mathrm{GJ} = 0.2$. We stress that in these conditions, we obtained an electrosphere configuration in simulations without pair production (see section~\ref{sec:global_sur}).

\begin{figure*}[t]
\centering
\includegraphics[width=4in]{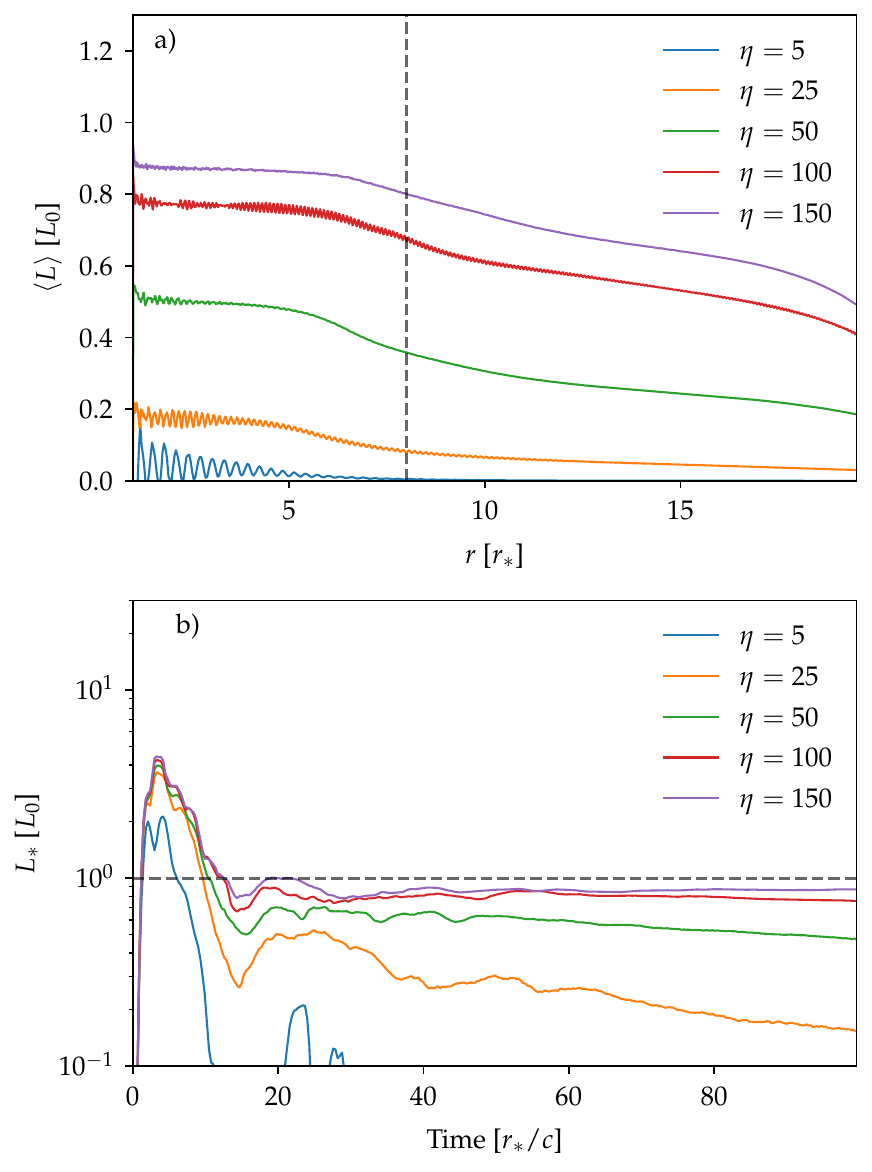}
\caption[Radial and temporal dependencies of Poynting flux in simulations with pair production.]{\label{fig:pp-L-avgs-comp}Radial and temporal dependencies of Poynting flux in simulations with pair production with varying $\eta$. a) shows the time-averaged luminosity $\langle L \rangle$ as a function of $r$ after a stellar rotation period, and b) shows the temporal evolution of the surface Poynting flux $L_*$. The dashed lines in a) and b) identify the light-cylinder radius and the theoretical surface Poynting flux $L_0 = \mu^2 \Omega^4 / c^3$, respectively.}
\end{figure*}

In Figure~\ref{fig:pp-rho-comp}, we show the charge density obtained at a time $t / T \simeq 2$ for a relevant subset of the simulations performed. We observe a transition from electrosphere to force-free-like configurations by increasing $\eta$. Physically, this corresponds to allowing \textcolor{coralpink}{more pairs per particle to be created}, hence increasing the plasma supply of the system. For $\eta = 5$, pair production is not efficient enough, and after an initial transient with some pair production, the accelerating electric field is screened and the system settles to an inactive solution. For $\eta \sim 10 - 50$, the system is able to launch plasma towards the light-cylinder and produce a positively charged equatorial outflow. This plasma is launched along the return current layers due to pair production at $r / r_* < 3$; however, because of the limited effectiveness of the pair production in this range of $\eta$, the plasma produced is not dense enough to confine the equatorial current sheet to a thin region, and it becomes wide for large distances from the stellar surface. For $\eta \gtrsim 100$, the system converges to a near force-free regime, with magnetic field lines open to infinity and a thin equatorial current sheet. In these simulations, pair production is very effective, and launches a large density ($n \sim$~few~$n_\mathrm{GJ}$), quasi-neutral plasma to the light-cylinder. In this region, part of the plasma escapes along the equatorial field lines; however, a fraction of the particles flows back to the star. The majority of these particles are electrons, such that the return current layers are negatively charged.

The time-averaged radial dependence of the Poynting flux and its surface value as a function of time for the simulations described above are presented in Figure~\ref{fig:pp-L-avgs-comp}. The observed radial dependence is similar to the regimes previously observed, with the $\eta \gtrsim 100$ simulations approaching the force-free spindown luminosity $L_0$ within the light-cylinder. In the equatorial current sheet, a fraction of $0.3-0.4~L_*$ is dissipated between $r \sim R_\mathrm{LC}$ and $r \sim 2~R_\mathrm{LC}$ and converted into particle kinetic energy. For all $\eta < 100$ runs, the surface luminosity decreases over time, and we expect them to eventually converge to the electrosphere solution for $t / T \gg 1$. However, for $\eta \gtrsim 100$, the surface Poynting flux remains stable over time.

\begin{figure*}[t]
\centering
\includegraphics[width=5.5in]{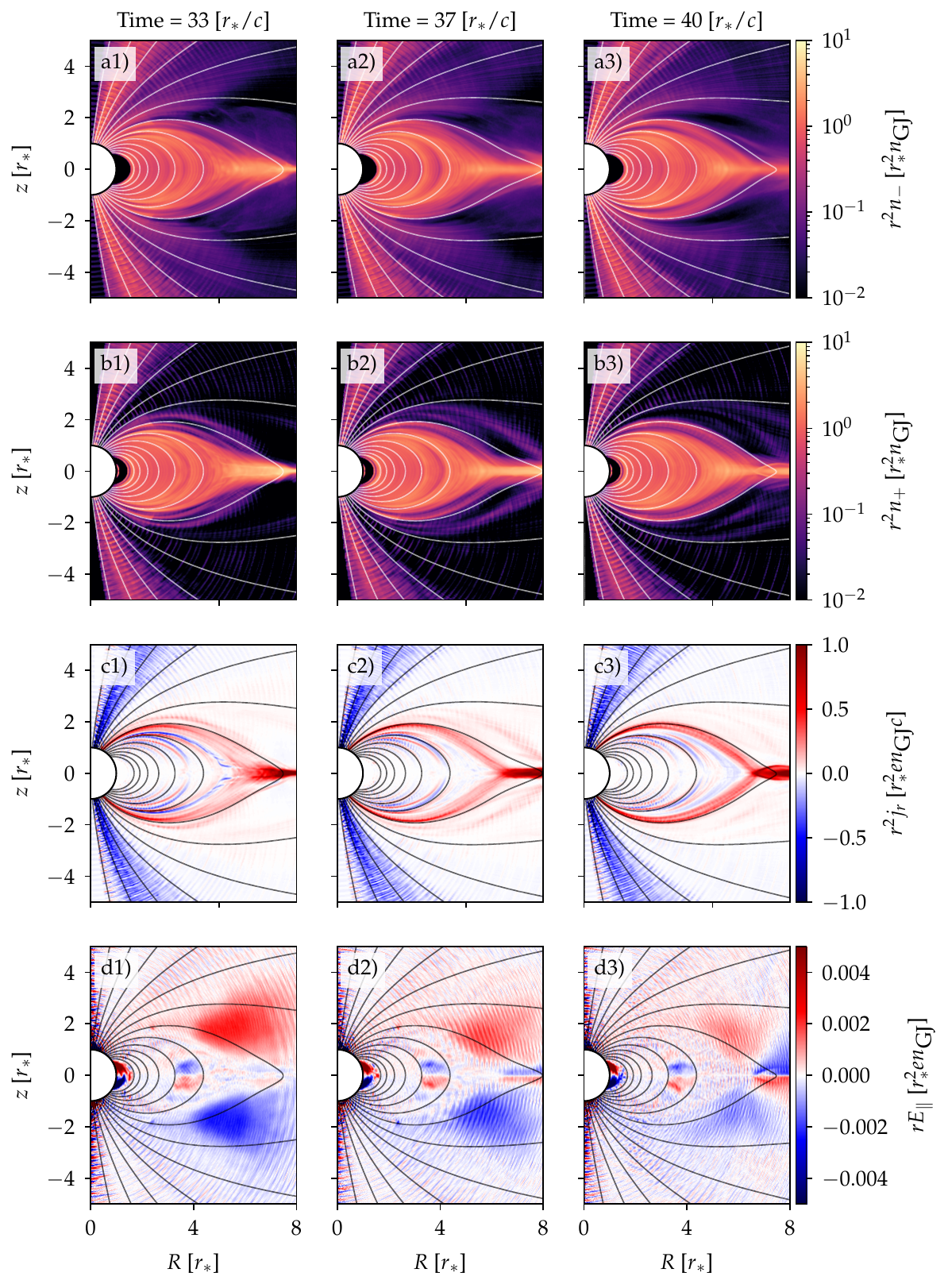}
\caption[Cyclic pair production along the return current layers.]{\label{fig:pp-cycle} Cyclic pair production along the return current layers. Columns labelled 1-3 correspond to different times and rows labelled a-d show the electron and positron densities, radial current and $E_\parallel$, respectively. Results obtained for $\eta = 100$. Solid lines represent magnetic field lines.}
\end{figure*}

All simulations present some temporal variability. We see small scale fluctuations on the charge and current densities in the open field line outflows, due to the $E_\parallel$ screening process resulting from pair cascades. These fluctuations occur on a temporal scale $\sim r_* / \eta c$. We also observe a quasi-periodic launch of plasma towards the light-cylinder region along the return current layers with a temporal scale $\sim 0.3 - 0.5~T$. We show one of these events in Figure~\ref{fig:pp-cycle} for a simulation with $\eta = 100$. As plasma is injected along the last closed field lines, most of it escapes along the equatorial current sheet. As this happens, the return current density drops close to $r \sim R_\mathrm{LC}$, allowing $E_\parallel$ to grow. Electrons flowing back to the star are thus accelerated along these field lines and produce a large number of pairs when they enter the pair producing region $r / r_* < 3$ --- see e.g., Figure~\ref{fig:pp-cycle} a1) and b1). The secondary particles then advect to large radii along the return current layers, reestablishing $j_r$ and effectively screening the $E_\parallel$ responsible for triggering the process --- see Figure~\ref{fig:pp-cycle} d1-3). This process produces a larger fraction of the total pair production events for $10 \lesssim \eta \lesssim 50$. The solutions obtained in this range resemble that of \textit{weak pulsars}~\citep{gruzinov_2015}, with screened surface $E_\parallel$ but with wide equatorial current sheets as a result of inefficient pair production. The process presented here is similar to that described in~\citet{chen_2020a, bransgrove_2022}.

The periodicity of the cyclic behaviour driven by pair production along the return current layers is $\sim 0.3 - 0.5~T$. We believe that this periodicity can depend on the multiplicity from the pair cascade near $r / r_* \sim 3$, since if more pairs outflow during the active phase, more electrons can be stored in the Y-point charge cloud, which takes longer to deplete. If this is true, a larger multiplicity should translate to a longer duty cycle. A detailed study of the importance of the cascade multiplicity on the cyclic behaviour is deferred to a future work.

\begin{figure*}[t]
\centering
\includegraphics[width=4.6in]{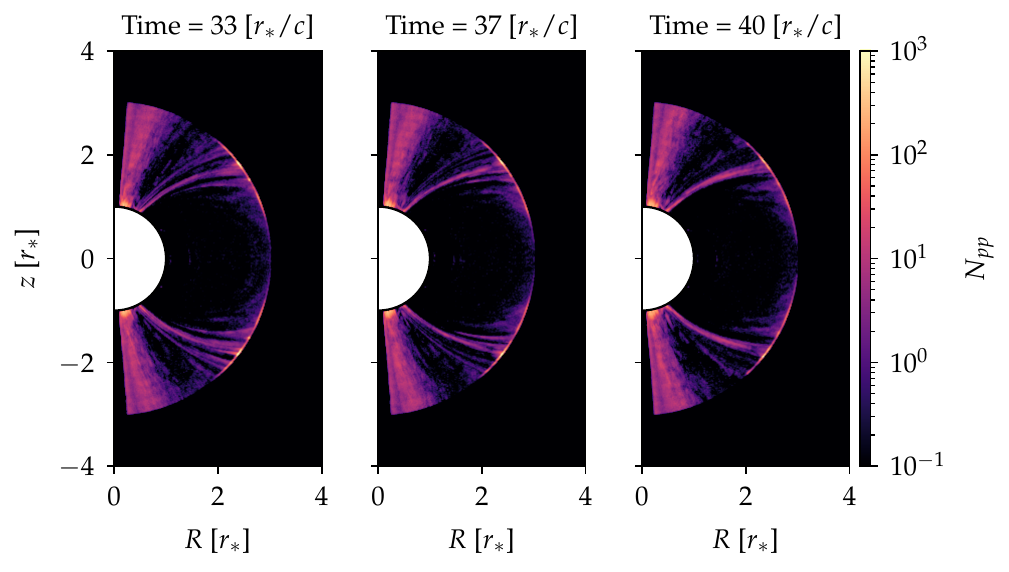}
\caption[Pair production sites.]{\label{fig:pp-loc}Pair production sites for the same simulation and times shown in Figure~\ref{fig:pp-cycle}. The color indicates the number of pair production events between data dumps ($\simeq 0.3~r_*/c$) in each grid cell.}
\end{figure*}

Finally, we note that apart from the effective pair discharges along the return current layers, we also observe abundant pair production within the polar cap region for all simulations with $\eta > 5$ --- see Figure~\ref{fig:pp-loc} for an illustrative example. This occurs because the density supplied from the stellar surface is insufficient to screen $E_\parallel$ in this region. With stronger surface injection, we expect this pair production to be less significant. However, we do not expect the overall structure of the magnetosphere to be meaningfully modified. Interestingly, the polar cap pair production observed in this regime resembles that expected when general relativity effects are taken into account. When corrections due to the strong gravitational field of the neutron star are considered, we expect pair creation activity within the polar cap even if the surface can supply a charge density $\pm e n_\mathrm{GJ}$~\citep{philippov_2015b, chen_2020a, bransgrove_2022}, since general relativity requires a current in this region $|j_r| > e n_\mathrm{GJ}$~\citep{beloborodov_2008, belyaev_2016, gralla_2016, torres_2023}. Apart from driving this difference in the time-dependent nature of the polar cap, general relativity is not expected to play a significant role in the overall magnetospheric organization.  

\section{Conclusions}
\label{sec:conclusions}

In this work, we have presented a systematic study of the different global regimes of pulsar magnetospheres. Namely, we have performed simulations with three distinct plasma sources: in volume, from the stellar surface, and via pair production. Our results, presented in Sect.~\ref{sec:global}, show that all plasma sources produce near force-free solutions in the regime of large plasma supply. In the opposite regime, we obtain inactive electrosphere solutions with all sources. These results are in overall good agreement with other works considering independently volume~\citep{philippov_2014, belyaev_2015, kalapotharakos_2018, brambilla_2018} or surface injection schemes~\citep{cerutti_2015, hakobyan_2023} or with heuristic pair production models~\citep{chen_2014, philippov_2015a, philippov_2015b, chen_2020a, guepin_2020, bransgrove_2022}.

While volume and surface plasma injection serve as a means to efficiently fill the pulsar magnetosphere and produce a near force-free configuration, as shown in Sects.~\ref{sec:global_vol} and \ref{sec:global_sur}, respectively, these are hard to motivate from first-principle arguments. On one hand, the pair cascades that these injection schemes aim to mimic develop only when the local magnetic field is close to the Schwinger field, and as such they should only operate near the stellar surface. On the other hand, these cascades produce plasma with a complex energy distribution, that depends on e.g., the local electric and magnetic field geometry. Thus, any volume or surface injection scheme is a substantial simplification of the highly nonlinear plasma supply from pair cascades in pulsars. Understanding if and how pair production alone can fill the whole pulsar magnetosphere is thus crucial, namely to reliably determine observational signatures.

The simulations including pair production presented in Sect.~\ref{sec:global_pp} show that pair discharges operating close to the stellar surface produce a range of solutions of the pulsar magnetosphere. The character of the solution depends critically on the ratio between the maximum attainable particle energy and the energy at which leptons emit pair producing photons, $\eta = \gamma_\mathrm{max} / \gamma_\mathrm{thr}$, that quantifies the efficiency of the pair discharges. Our results show that when $\eta \gtrsim 100$, enough pairs are created to fill the magnetosphere and reach a near force-free surface Poynting flux, with dissipation occurring in an equatorial current sheet beyond the light-cylinder. In the opposite limit, $\eta \lesssim 10$, the magnetosphere settles to a fully charge-separated, static solution, with $E_\parallel = 0$ near the surface, that produces a negligible Poynting flux. For $\eta \sim 10 - 50$, we observe an intermediate solution~\citep{gruzinov_2015}, with a wide equatorial current sheet and with a surface Poynting flux $50 - 80\%$ below that expected in the force-free regime.

Our simulations show that the pair production along the return current layers is key to feed plasma to the light-cylinder region and beyond in near force-free regimes\textcolor{coralpink}{, in line with the results reported in other works, e.g.~\citet{chen_2014}}. We have also identified a time-dependent mechanism \textcolor{coralpink}{similar to that presented in~\citet{chen_2020a, bransgrove_2022}}, that results from periodic openings of an outer gap in which particles flowing back to the star are able to accelerate, producing pairs when they get close to the stellar surface.

The simulations presented here used a very simple heuristic model to describe pair production in strong magnetic fields. In this work, we have only explored the role of the parameter $\eta$ on the magnetospheric structure and left the ratio $\gamma_\mathrm{thr} / \gamma_\mathrm{pair}$ unchanged. This ratio plays an important role in the multiplicity of pair cascades, and was kept low to make simulations feasible. Larger values of $\gamma_\mathrm{thr} / \gamma_\mathrm{pair}$ will likely provide even more abundant pairs to large radii, such that smaller values of $\eta$ may be enough to set the magnetosphere in a force-free regime. Such study is left for future work.

The pair production model considered here provides an adequate description of pair cascades when the curvature photon mean free path is negligible, \textit{i.e.}, when pair production is local. In global models, however, it is easy to conceive that photons emitted in some regions of the magnetosphere may decay into pairs in others. For instance, photons emitted by electrons travelling towards the star along the return current layer may decay in the polar cap region. It would thus be interesting to include more sophisticated pair production models in these simulations to assess if nonlocal pair production may play a significant role in e.g., coherent emission processes.

In this work, we have also described a spherical grid suitable to perform global PIC simulations of pulsar magnetospheres. We have detailed a) an electromagnetic field solver based on the Yee solver that uses an integral form of Maxwell's equations (Sect.~\ref{sec:spherical_solver}, b) particle pushers that solve the particles' equations of motion in Cartesian coordinates (Sect.~\ref{sec:spherical_pusher}) and c) a charge-conserving current deposition scheme (Sect.~\ref{sec:spherical_current}) for a non-uniform, curvilinear spherical grid. While the field solver and particle pusher techniques are also implemented in other similar codes, the current deposition scheme presented here is a novel development. By ensuring that the continuity equation (and, consequently, Gauss' law) is satisfied in the current deposition, this method does not require that other numerical algorithms are used to correct for artificial charges in the grid. For each of the numerical schemes presented here, we have provided comprehensive benchmarks for a variety of test scenarios. All numerical schemes presented here have been implemented in the PIC code OSIRIS.

\section{Acknowledgments}
FC, TG, RAF and LOS acknowledge supported by the European Research Council (ERC-2015-AdG Grant 695088) and FCT (Portugal)— Foundation for Science and Technology (grant PD/BD/114307/2016, in the framework of the Advanced Program in Plasma Science and Engineering APPLAuSE, grant PD/00505/2012, and project no. 2022.02230.PTDC). AC acknowledges support from NSF grants DMS-2235457 and AST-2308111. AS is supported in part by NSF grant PHY-2206607. We acknowledge PRACE for granting access to MareNostrum, Barcelona Supercomputing Center (Spain), where the simulations presented in this work were performed.

\bibliographystyle{aa}
\bibliography{aanda}

\end{document}